\documentclass[journal,10pt,onecolumn, draftclsnofoot, ]{IEEEtran}
\usepackage{setspace}
\usepackage{cite}
\usepackage{amsmath,amssymb,amsfonts}

\usepackage{algorithmic}
\usepackage{algorithm}
\usepackage{graphicx}
\usepackage{textcomp}
\usepackage{xcolor}
\usepackage{amsthm}

\usepackage[center]{caption} 
\usepackage{subcaption}

\def\BibTeX{{\rm B\kern-.05em{\sc i\kern-.025em b}\kern-.08em
		T\kern-.1667em\lower.7ex\hbox{E}\kern-.125emX}}
\begin{document}
	
	\title{Tracking an Underwater Target with Unknown Measurement Noise Statistics Using Variational Bayesian Filters\\
	}
		
	\author{\IEEEauthorblockN{Shreya Das}\\
		\IEEEauthorblockA{\textit{Dept. of Elect. Engg.} \\
			\textit{IIT Patna}\\
			Patna, India \\
			shreya\_2121ee15@iitp.ac.in}\\
		\and
		\IEEEauthorblockN{Kundan Kumar}\\
		\IEEEauthorblockA{\textit{Dept. of Elect. Engg. and Automation} \\
			\textit{Aalto University,}\\
			Finland \\
			kundan.kumar@aalto.fi}\\
		\and
		\IEEEauthorblockN{Shovan Bhaumik}\\
		\IEEEauthorblockA{\textit{Dept. of Elect. Engg.} \\
			\textit{IIT Patna}\\
			Patna, India \\
			shovan.bhaumik@iitp.ac.in}	
	}

	\maketitle
	
	\begin{abstract}
		This paper considers a bearings-only tracking problem using noisy measurements of unknown noise statistics from a passive sensor. It is assumed that the process and measurement noise follows the Gaussian distribution where the measurement noise has an unknown non-zero mean and unknown covariance. Here an adaptive nonlinear filtering technique is proposed where the joint distribution of the measurement noise mean and its covariance are considered to be following normal inverse Wishart distribution (NIW). Using the variational Bayesian (VB) method the estimation technique is derived with optimized tuning parameters \emph{i.e.,} the confidence parameter and the initial degree of freedom of the measurement noise mean and the covariance, respectively. The proposed filtering technique is compared with the adaptive filtering techniques based on maximum likelihood and maximum aposteriori in terms of root mean square error in position and velocity, bias norm, average normalized estimation error squared, percentage of track loss, and relative execution time.
		Both adaptive filtering techniques are implemented using the traditional Gaussian approximate filters and are applied to a bearings-only tracking problem illustrated with moderately nonlinear and highly nonlinear scenarios to track a target following a nearly straight line path. Two cases are considered for each scenario, one when the measurement noise covariance is static and another when the measurement noise covariance is varying linearly with the distance between the target and the ownship. In this work, the proposed adaptive filters using the VB approach are found to be superior to their corresponding adaptive filters based on the maximum aposteriori and the maximum likelihood at the expense of higher computation cost.
	\end{abstract}
	
	
	
	
	\section{Introduction}
	Target motion analysis (TMA) is the problem of tracking the kinematics of a target, \emph{i.e.} localizing the target and estimating its velocity using noisy measurements. When measurements are obtained from a single observer the problem is termed autonomous TMA \cite{ristic2003beyond, leong2013, radhakrishnan2018}. In TMA, the use of only the line of sight (LOS) angle of the target \emph{i.e.} the bearing angle as a measurement to estimate the kinematics of a target is commonly referred to as the bearings-only tracking (BOT) problem. The BOT finds several challenging applications in the practical world having major significance in defense applications like for example tracking of submarine \cite{8009822, annabattula2015multi}, torpedo \cite{kumar2022tracking, karlsson2005recursive} using bearing measurements obtained from passive sonars and aircraft surveillance using passive bearing measurements from radar \cite{farina1999target}. A vital reason for using these passive bearing measurements in defense applications is to prevent revealing the ownship's orientation to the enemies, especially during times of conflict \cite{radhakrishnan2015quadrature}. In such a case, an outstanding performance of BOT becomes crucial.
	
	So BOT is extensively explored in literature and it has been observed that the dynamic model of autonomous TMA becomes unobservable if the observer does not maneuver and follow a constant velocity model while tracking a nonmaneuvering target \cite{nardone1981observability}. In order to make the system observable, the observer has to outmaneuver the target \cite{song1999observability}. BOT for maneuvering target is also discussed in literature \cite{ristic2003tracking, arulampalam2004bearings, li2014bearings, kirubarajan2001bearings}.
	
	
	Due to the nonlinearity present in the bearing measurements, the optimal Kalman filter becomes ineffective in estimating the location and the velocity of the target as the optimal solution leads to mathematically intractable integrals \cite{4102101, aidala1983utilization}. Consequently, we have to settle with the suboptimal nonlinear filtering algorithms, the first and the most popular among which is the extended Kalman filter (EKF). The EKF \cite{toloei2014state} and its variants \cite{song1985stochastic, rao2005modified, jawahar2016modified} deal with the nonlinearities simply by linearizing them using the first order Taylor series approximation. But the EKF suffered from drawbacks as it often resulted in unstable performances that included poor estimation accuracy and high track divergence \cite{4102101, nardone1984fundamental}. To compensate for these disadvantages of EKF several other nonlinear filtering algorithms were developed namely the cubature Kalman filter (CKF) \cite{arasaratnam2009cubature}, the unscented Kalman filter (UKF) \cite{julier2004unscented, xu2004single}, the Gauss-Hermite filter (GHF) \cite{ito2000gaussian, chalasani2012bearing} and their variants \cite{bhaumik2013cubature, xiong2009modified, radhakrishnan2016multiple}. These filters approximate the probability density functions using deterministic sample points and their corresponding weights. The process and the measurement equations update the sample points and the weights. 
	
	All the above mentioned Gaussian approximated filters (GAFs) consider the measurement noise to be Gaussian in nature and their performance is greatly affected by the prior knowledge of the measurement noise statistics. These filters perform best when the measurement noise statistics \emph{i.e.} the mean and the covariance of the measurement noise are known. But in most practical applications, the statistics of the measurement noise remain unknown. In such a situation the estimation performance of these filters degrades. 
	
	Lack of information on the measurement noise statistics may lead to biased estimation or track divergence. If the covariance of the measurement noise is assumed to be very less than the truth then the estimation becomes biased while if it is very high compared to the true value then the filter loses track \cite{mohamed1999adaptive}. So estimation of the measurement noise statistics directly affects the filtering performance. So an adaptive UKF using the maximum a posteriori (MAP) method is proposed in \cite{zhao2009adaptive} to estimate the non-zero mean as well as the covariance of both the process and measurement noise along with the states of the target. In a similar approach, an adaptive square root cubature Kalman filter was proposed in \cite{zhao2015design} to enhance the adaptivity and robustness of the square root Kalman filter. But the technique cannot guarantee the convergence to the right noise covariance and the positive definiteness of the estimated noise covariance. Due to its numerical instability and filtering divergence, this process of adaptation is not always favorable.
	
	Another adaptive filtering technique was developed \cite{mehra1995adaptive, lee2003centralized, jwo2009adaptive}. R-adaptive filter was developed using the maximum likelihood estimation (MLE) of the innovation based sequence and the residual based sequence \cite{almagbile2010evaluating}. The residual based R-adaptive UKF and the innovation based R-adaptive UKF was proposed in \cite{das2013adaptive} which was analogous to the R-adaptive Kalman filter in \cite{almagbile2010evaluating}. Using the estimated R the process noise covariance, Q is also estimated. The positive definiteness of the estimated noise covariance matrix is guaranteed in the residual adaptive method based on MLE. However, these papers consider the process and measurement noise to be Gaussian having a known zero mean. 
	
	Recently, an adaptive technique to estimate the target state with the process and measurement noise having unknown covariance and known zero mean using inverse Wishart (IW) distribution for linear Kalman filter was proposed in \cite{huang2020variational}. The same technique implemented for nonlinear extended Kalman filter (EKF) was proposed in \cite{sun2018new} and for cubature Kalman filter in \cite{zhang2020robust}. Here, the measurement noise covariance matrix and the predicted error covariance matrix are approximated by inverse Wishart distribution. This technique was further explored to estimate the target state with inaccurate process and measurement noise non zero mean and covariance for the linear Kalman filter in \cite{xu2019new}. 
	
	In our work, we developed an adaptive filtering technique to estimate the target state for BOT problems when the measurement noise non zero mean and covariance are unknown using VB based technique, optimizing the likelihood to find the prior confidence parameter of measurement mean and degree of freedom of the measurement covariance, for nonlinear Gaussian filters. The results of it are compared with the MAPMLE based adaptive technique. In the MAPMLE based technique, we estimated the measurement noise mean using the MAP method and the covariance using the MLE method to ensure the positive definiteness of the estimated error covariance matrix.
	
	The adaptive methods were implemented in two scenarios, one moderately nonlinear and the other highly nonlinear to track a target moving with constant velocity, with both static and varying measurement noise covariance. The performance of adaptive filters using the VB method is compared with the adaptive filters with the MAPMLE method and the corresponding nonadaptive filters with known measurement noise statistics in terms of root mean square error (RMSE) in position and velocity, bias norm, average normalized estimation error squared (ANEES), percentage of track divergence, and relative execution time. In both scenarios, the adaptive filters using the VB approach have better accuracy than the adaptive filters using the MAPMLE methods at the expense of more execution time. 
	
	The organization of the paper is as follows: Section II contains the problem formulation that included the formulation of the process model and the measurement model. Section III describes the normal inverse Wishart distribution. Section IV derives the VB method of adaptive filtering. Section V provides a brief explanation of the MAPMLE technique of adaptive filters. Section VI deals with the properties of both the adaptive filtering approaches used in the paper. Section VII provides the simulation results. The paper ends with a brief discussion and conclusion in Section VIII.

	\section{Problem Formulation}
	\subsection{System model}
	The state space dynamic of the target that is assumed to follow a nearly straight line path with constant velocity in the discrete time domain is expressed as $\mathcal{X}_k^{t}=\begin{bmatrix}
		x_k^{t} & y_k^{t} & \dot{x}_k^{t} & \dot{y}_k^{t}
	\end{bmatrix}$, where $\begin{bmatrix}
		x_k^{t} & y_k^{t}
	\end{bmatrix}$ and $\begin{bmatrix}
		\dot{x}_k^{t} & \dot{y}_k^{t}
	\end{bmatrix}$ are the target's position and velocity at the $k$-th time instant, respectively. The same way follows the observer state space dynamic which can be expressed as $\mathcal{X}_k^{o}=\begin{bmatrix}
		x_k^{o} & y_k^{o} & \dot{x}_k^{o} & \dot{y}_k^{o}
	\end{bmatrix}$, where $\begin{bmatrix}
		x_k^{o} & y_k^{o}
	\end{bmatrix}$ and $\begin{bmatrix}
		\dot{x}_k^{o} & \dot{y}_k^{o}
	\end{bmatrix}$ are the observer's position and velocity at the $k$-th time instant, respectively.
	The dynamic model of the target can be expressed as \cite{ristic2003beyond, leong2013}
	\begin{align}
		\mathcal{X}_{k}= F\mathcal{X}_{k-1}+\omega_{k-1}- \mho_{k-1,k},  \label{process_model}
	\end{align}
	where $\mathcal{X}_{k}=\mathcal{X}_k^{t}-\mathcal{X}_k^{o}=
	\begin{bmatrix}
		x_{k} & y_{k} & \dot x_{k} & \dot y_{k}
	\end{bmatrix}^T$
	is the relative state vector between target and observer, $F$ is  the state transition matrix evaluated as:
	\begin{equation}
		F=\begin{bmatrix}
			1 & 0 & \Delta & 0\\
			0 & 1 & 0 & \Delta\\
			0 & 0 & 1 & 0 \\
			0 & 0 & 0 & 1
		\end{bmatrix},
	\end{equation} where $\Delta$ is the sampling time. $\mho_{k-1,k}$  is a vector of inputs evaluated as:
	\begin{equation}
		\mho_{k-1,k}=\begin{bmatrix}
			x_k^{o}-x_{k-1}^{o}-\Delta\dot{x}_{k-1}^{o}\\
			y_k^{o}-y_{k-1}^{o}-\Delta\dot{y}_{k-1}^{o}\\
			\dot{x}_k^{o}-\dot{x}_{k-1}^{o}\\
			\dot{y}_k^{o}-\dot{y}_{k-1}^{o}
		\end{bmatrix},
	\end{equation} and $\omega_{k-1} \sim \mathcal{N}(0, Q_{k-1})$ is the process noise with $q_m$ mean and covariance $Q_{k-1}$ given as:
	\begin{equation}
		Q=\begin{bmatrix}
			\dfrac{\Delta^3}{3} & 0 & \dfrac{\Delta^2}{2} & 0\\
			0 & \dfrac{\Delta^3}{3} & 0 & \dfrac{\Delta^2}{2}\\
			\dfrac{\Delta^2}{2} & 0 & \Delta & 0\\
			0 & \dfrac{\Delta^2}{2} & 0 & \Delta
		\end{bmatrix}\bar{q},
	\end{equation} where $\bar{q}$ is the intensity of process noise.	
	\subsection{Measurement model}
	The bearing measurement of the target with respect to true north is measured using sensors mounted to own ship. The measurement model is expressed as:
	\begin{equation}\label{meas_BOT}
		\mathcal{Y}_k = h(\mathcal{X}_k) + \nu_{\theta_k},
	\end{equation}
	where $h(\mathcal{X}_k)=\text{tan}^{-1}(x_k/y_k)$ and $\nu_{\theta_k}$ is the measurement noise with $r_m$ mean and $\sigma_{\theta_k}$ standard deviation \emph{i.e.} $\nu_{\theta_k} \sim \mathcal{N}(r_m, \sigma_{\theta_k}^2)$. To make the system observable, the the own ship has to maneuver to estimate the target state \cite{ristic2003beyond, leong2013}.  
	
	
	
	\section{Normal Inverse Wishart (NIW) Distribution}


	
	The inverse Wishart distribution of a random symmetric positive definite matrix, $\mathcal{B} \in \mathbb{R}^{n \times n}$ can be written as
	\begin{equation}
		IW(\mathcal{B};\lambda,\psi)=\dfrac{|\psi|^{\lambda/2}|\mathcal{B}|^{-(\lambda+n+1)/2}\exp{\{-0.5tr(\psi \mathcal{B}^{-1})\}}}{2^{n\lambda/2}\Gamma_n(\lambda/2)},
	\end{equation} 
	where $\lambda$ is the degrees of freedom, $\psi  \in \mathbb{R}^{n \times n}$ is a symmetric positive definite matrix and is known as the scale matrix,  $|\cdot|$, $tr(\cdot)$ are the determinant and trace of a matrix, respectively, and $\Gamma_n(\cdot)$ is the $n$ variate gamma function \cite{o2004kendall}.  If $\mathcal{B}\sim IW(\mathcal{B}; \lambda, \psi)$, then a property of the IW distribution is that the mean of the distribution \emph{i.e.} $E[\mathcal{B}]=\psi/(\lambda-n-1)$ such that $\lambda>n+1$ \cite{o2004kendall}. 
	
	

	The NIW probability density function of $\mathcal{X}$ and $\mathcal{B}$ can be expressed as
	\begin{equation}\label{NIWpdfJoint}
		NIW(\mathcal{X},\mathcal{B};\mu_\mathcal{X},\alpha,\lambda,\psi)=\mathcal{N}\left(\mathcal{X};\mu_\mathcal{X},\alpha\mathcal{B}\right) IW(\mathcal{B};\lambda,\psi),
	\end{equation}
	where $\mu_\mathcal{X}=\mathbb{E}[\mathcal{X}]$ and $\alpha>0$ is named as the confidence parameter. With marginalization the pdf of a NIW distributed random vector $\mathcal{X}$ can be written as 
	\begin{equation}\label{NIWpdf}
		NIW(\mathcal{X};\mu_\mathcal{X},\alpha,\lambda,\psi)=\int\mathcal{N}(\mathcal{X};\mu_\mathcal{X},\alpha\mathcal
		B)\times IW(\mathcal{B};\lambda,\psi)d\mathcal{B}.
	\end{equation}

	When the normal distribution mean is known, the inverse Wishart (IW) distribution is used as the conjugate prior for covariance of normal distribution \cite{o2004kendall}. But in case when the mean as well as the covariance of a normal distribution is unknown then in general, the normal inverse Wishart \emph{i.e.} NIW distribution is preferred as the joint conjugate prior for mean and covariance of a Gaussian distribution in Bayesian statistics \cite{murphy2007conjugate}.

	\section{Variational Bayesian Technique of Adaptive Filtering}
	\subsection{Process update}
	The prior state follows Gaussian distribution with mean, $\hat{\mathcal{X}}_{k|k-1}$ and covariance, $P_{k|k-1}$ \emph{i.e.},
	\begin{equation}
		p(\mathcal{X}_{k|k-1}|\mathcal{Y}_{1:k-1})=\mathcal{N}(\mathcal{X}_{k|k-1},\hat{\mathcal{X}}_{k|k-1},P_{k|k-1}),
	\end{equation}
	As the process noise statistics are known \emph{i.e.,} $\mathcal{N}(0,Q_k)$ and the process is linear in our case so the process update can be performed as follows,
	\begin{equation}
		\hat{\mathcal{X}}_{k|k-1}=F\hat{\mathcal{X}}_{k-1|k-1}-\mho_{k-1,k},
	\end{equation}
	\begin{equation}
		P_{k|k-1}=FP_{k-1|k-1}F^T+Q_{k-1}.
	\end{equation}
	\subsection{Measurement update}
	As the measurement noise statistics are unknown so we start off with a guess value for both the mean and the covariance. Considering the guess value of measurement noise mean and covariance be denoted by $r'_m$ and $R'_k$, respectively. \\
	\textbf{Lemma 1:} The joint pdf of $p(\mathcal{X}_{k|k-1},R'_k,r'_m,\mathcal{Y}_{1:k})$ can be factored as
	\begin{equation}\label{Eq_mulprob}
		p(\mathcal{X}_{k|k-1},R'_k,r'_m,\mathcal{Y}_{1:k})=p(\mathcal{Y}_k|\mathcal{X}_{k|k-1},R'_k,r'_m)p(\mathcal{X}_{k|k-1}|\mathcal{Y}_{1:k-1})p(r'_m|R'_k,\mathcal{Y}_{1:k-1})p(R'_k|\mathcal{Y}_{1:k-1})p(\mathcal{Y}_{1:k-1}).
	\end{equation}
	\begin{proof}
		The joint pdf of $p(\mathcal{X}_{k|k-1},R'_k,r'_m,\mathcal{Y}_{1:k})$ can be decomposed using Bayes' theorem as follows,
		\begin{equation*}\label{Eq_FactorPDF}
			\begin{split}
				p(\mathcal{X}_{k|k-1},R'_k,r'_m,\mathcal{Y}_{1:k})&=p(\mathcal{X}_{k|k-1}, R'_{k}, r'_m\mathcal{Y}_{k},\mathcal{Y}_{1:k-1})\\&
				=p(\mathcal{Y}_k|\mathcal{X}_{k|k-1},R'_k,r'_m,\mathcal{Y}_{1:k-1})p(\mathcal{X}_{k|k-1},R'_k,r'_m,\mathcal{Y}_{1:k-1})\\&
				=p(\mathcal{Y}_k|\mathcal{X}_{k|k-1},R'_k,r'_m)p(\mathcal{X}_{k|k-1}|R'_k,r'_m,\mathcal{Y}_{1:k-1})p(R'_k,r'_m,\mathcal{Y}_{1:k-1})\\&
				=p(\mathcal{Y}_k|\mathcal{X}_{k|k-1},R'_k,r'_m)p(\mathcal{X}_{k|k-1}|\mathcal{Y}_{1:k-1})p(r'_m|R'_k,\mathcal{Y}_{1:k-1})p(R'_k,\mathcal{Y}_{1:k-1})\\&
				=p(\mathcal{Y}_k|\mathcal{X}_{k|k-1},R'_k,r'_m)p(\mathcal{X}_{k|k-1}|\mathcal{Y}_{1:k-1})p(r'_m|R'_k,\mathcal{Y}_{1:k-1})p(R'_k|\mathcal{Y}_{1:k-1})p(\mathcal{Y}_{1:k-1}).
			\end{split}
		\end{equation*} 
	\end{proof}
	
	As $R'_k$ is covariance of Gaussian pdf, we consider the prior distribution of $p(R'_k|\mathcal{Y}_{1:k-1})$ follows IW distribution. So, we can write
	\begin{equation}\label{Eq_R'IW}
		p(R'_k|\mathcal{Y}_{1:k-1})=IW(R'_k;\hat{u}'_{k|k-1},\hat{U}'_{k|k-1}).
	\end{equation}
	We can say from \eqref{Eq_R'IW}, $E[R'_k]=\dfrac{\hat{U}'_{k|k-1}}{\hat{u}'_{k|k-1}-m-1}$; we assume $\hat{u}'_{k|k-1}>m+1$. As $r'_m$ is the mean of Gaussian pdf, we consider the prior distribution of $p(r'_m|\mathcal{Y}_{1:k-1})$ follows normal distribution,
	\begin{equation}\label{Eq_r'N}
		p(r'_m|\mathcal{Y}_{1:k-1})=\mathcal{N}(r'_m;\mu',\alpha' R'_k).
	\end{equation}
	So, using inverse Wishart and normal distribution of \eqref{Eq_R'IW} and \eqref{Eq_r'N}, \eqref{Eq_mulprob} can be written as
	\begin{equation}
		\begin{split}
			p(\mathcal{X}_{k|k-1},R'_k,r'_m,\mathcal{Y}_{1:k})&=\mathcal{N}(\mathcal{Y}_k;h({\mathcal{X}}_{k|k-1})+r'_m,R'_k)\mathcal{N}(\mathcal{X}_{k|k-1};\hat{\mathcal{X}}_{k|k-1},P_{k|k-1})\mathcal{N}(r'_m;\mu',\alpha 'R'_k)\\&IW(R'_k;\hat{u}'_{k|k-1},\hat{U}'_{k|k-1}).
		\end{split}
	\end{equation}


	\textbf{Theorem 1:} 
	The joint pdf $p(\mathcal{X}_{k|k-1},r'_m,R'_k|\mathcal{Y}_{1:k})$ can be approximately factorized as 
	\begin{equation}
		p(\mathcal{X}_{k|k-1},r'_m,R'_k|\mathcal{Y}_{1:k})\approx q(\mathcal{X}_{k|k-1})q(r_m)q(R_k),
	\end{equation}
	where $q(\cdot)$ represents approximate pdf of $p(\cdot)$ and can be expressed as 
	\begin{equation}\label{Eq_Solution}
		q(\theta)=E_{-\theta}[\log p(\mathcal{X}_{k|k-1},r'_m,R'_k,\mathcal{Y}_{1:k})]+c_{\theta},
	\end{equation}
	where the $E_{-\theta}$ represents the expectation on elements other than $\theta$ and $c_{\theta}$ is a constant. 
	\begin{proof}
		By minimizing the Kullback-Leibler divergence (KLD) between the factored approximate pdf $q(\mathcal{X}_{k|k-1})q(r_m)q(R_k)$ and the true posterior pdf $p(\mathcal{X}_{k|k-1},r'_m,R'_k|\mathcal{Y}_{1:k})$ we will receive \eqref{Eq_Solution}. It has been proved in page 450 of \cite{bishop2006pattern} and (15) of \cite{tzikas2008variational}.
	\end{proof}
	\textbf{Theorem 2:} Approximate pdf of $q(\mathcal{X}_{k|k-1})$ can be expressed as
	\begin{equation}
		q(\mathcal{X}_{k|k-1})\sim\mathcal{N}(\mathcal{Y}_k;h(\mathcal{X}_{k|k-1})+r_m,R_k)\mathcal{N}(\mathcal{X}_{k|k-1};\hat{\mathcal{X}}_{k|k-1},P_{k|k-1}),
	\end{equation}
	where 
	\begin{equation}\label{Eq_R}
		R_k=\dfrac{\hat{U}'_{k|k-1}}{\hat{u}'_{k|k-1}-m-1}.
	\end{equation}
	
	\begin{proof}
		The proof is provided in Appendix A.
	\end{proof} 
	
	\textbf{Theorem 3:} Approximate pdf of $q(r_m)$ can be expressed as 
	\begin{equation}
		q(r_m)\sim \mathcal{N}(r_m; \mu ,\alpha R_k),
	\end{equation}
	where 
	\begin{equation}\label{Eq_mu}
		\mu=\dfrac{\mu'+\alpha'(\mathcal{Y}_k-\sum_{j=1}^{N_s}w_j\mathtt{Y}_{j,k|k-1})}{\alpha'+1},
	\end{equation}
	where $N_s$ is the number of sigma points, $\xi_j$ and $w_j$ are the $j$th sigma point and its corresponding weight, respectively, $\mathtt{Y}_{j,k|k-1}=h(\mathtt{X}_{j,k|k-1})=h(S_{k|k-1}\xi_j+\hat{\mathcal{X}}_{k|k-1})$, where $S_{k|k-1}S_{k|k-1}^T=P_{k|k-1}$, and 
	\begin{equation}\label{Eq_alpha}
		\alpha =\dfrac{\alpha'}{\alpha'+1}.
	\end{equation}
	
	\begin{proof}
		The proof is provided in Appendix B.
	\end{proof} 
	
	\textbf{Theorem 4:} Approximate pdf of $q(R_k)$ can be expressed as 
	\begin{equation}
		q(R_k)\sim IW(R_k;\hat{u}_{k|k-1},\hat{U}_{k|k-1}),
	\end{equation}
	such that 
	\begin{equation}\label{Eq_u}
		\hat{u}_{k|k-1}=\hat{u}'_{k|k-1}+2,
	\end{equation}
	and 
	\begin{equation}\label{Eq_U}
		\hat{U}_{k|k-1}=\hat{U}'_{k|k-1}+B_k+D_k,
	\end{equation}
	where 
	\begin{equation}\label{Eq_BkFirst}
		B_k=E_{\mathcal{X}_{k|k-1},r'_m} [(\mathcal{Y}_{k}-h(\mathcal{X}_{k|k-1})-r'_m)(\mathcal{Y}_{k}-h(\mathcal{X}_{k|k-1})-r'_m)^T]
	\end{equation}
	and 
	\begin{equation}\label{Eq_DkFirst}
		D_k=\dfrac{1}{\alpha'}E_{r'_m}[(r'_m-\mu')(r'_m-\mu')^T]
	\end{equation}. 
	
	\begin{proof}
		The proof is provided in Appendix C.
	\end{proof} 
	
	\textbf{Lemma 2:} $B_k$ of \eqref{Eq_BkFirst} can be manipulated as 
	\begin{equation}\label{Eq_Bk}
		B_k=(\mathcal{Y}_{k}-E[h(\mathcal{X}_{k|k-1})]-\mu)(\mathcal{Y}_{k}-E[h(\mathcal{X}_{k|k-1})]-\mu)^T+\sum_{j=1}^{N_s}w_j[\mathtt{Y}_{j,k|k-1}-E[h(\mathcal{X}_{k|k-1})]][\mathtt{Y}_{j,k|k-1}-E[h(\mathcal{X}_{k|k-1})]]^T+\alpha' R'_{k},
	\end{equation}
	where $E[h(\mathcal{X}_{k|k-1})]=\sum_{j=1}^{N_s}w_j\mathtt{Y}_{j,k|k-1}$.
	
	\begin{proof}
		Evaluating $B_k$,
		\begin{equation}
			\begin{split}
				B_k&=E_{\mathcal{X}_{k|k-1},r'_m} [(\mathcal{Y}_{k}-h(\mathcal{X}_{k|k-1})-r'_m)(\mathcal{Y}_{k}-h(\mathcal{X}_{k|k-1})-r'_m)^T]\\&
				=E_{\mathcal{X}_{k|k-1},r'_m} [(\mathcal{Y}_{k}-E[h(\mathcal{X}_{k|k-1})]+E[h(\mathcal{X}_{k|k-1})]-\mu'+\mu'-h({\mathcal{X}}_{k|k-1})-r'_m)(\mathcal{Y}_{k}-E[h(\mathcal{X}_{k|k-1})]\\&+E[h(\mathcal{X}_{k|k-1})]-\mu'+\mu'-h({\mathcal{X}}_{k|k-1})-r'_m)^T]\\&
				=(\mathcal{Y}_{k}-E[h(\mathcal{X}_{k|k-1})]-\mu')(\mathcal{Y}_{k}-E[h(\mathcal{X}_{k|k-1})]-\mu')^T+(\mathcal{Y}_k-E[h(\mathcal{X}_{k|k-1})]-\mu')(E[h(\mathcal{X}_{k|k-1})]]\\&-E[h(\mathcal{X}_{k|k-1})]+\mu'-E[r'_m])^T+(E[h(\mathcal{X}_{k|k-1})]]-E[h(\mathcal{X}_{k|k-1})]+\mu'-E[r'_m])(\mathcal{Y}_k-E[h(\mathcal{X}_{k|k-1})]-\mu')^T\\&+E_{\mathcal{X}_{k|k-1},r'_m} [\{h(\mathcal{X}_{k|k-1})-E[h(\mathcal{X}_{k|k-1})]+\mu'-r'_m\}\{h(\mathcal{X}_{k|k-1})-E[h(\mathcal{X}_{k|k-1})]+\mu'-r'_m\}^T]\\&
				=(\mathcal{Y}_{k}-E[h(\mathcal{X}_{k|k-1})]-\mu')(\mathcal{Y}_{k}-E[h(\mathcal{X}_{k|k-1})]-\mu')^T+E_{\mathcal{X}_{k|k-1},r'_m} [\{h(\mathcal{X}_{k|k-1})-E[h(\mathcal{X}_{k|k-1})]+\mu'-r'_m\}\\&\{h(\mathcal{X}_{k|k-1})-E[h(\mathcal{X}_{k|k-1})]+\mu'-r'_m\}^T].
			\end{split}
		\end{equation}
		Now, 
		\begin{equation}
			\begin{split}
				&E_{\mathcal{X}_{k|k-1},r'_m} [\{h(\mathcal{X}_{k|k-1})-E[h(\mathcal{X}_{k|k-1})]+\mu'-r'_m\}\{h(\mathcal{X}_{k|k-1})-E[h(\mathcal{X}_{k|k-1})]+\mu'-r'_m\}^T]\\&=E[\{h(\mathcal{X}_{k|k-1})-E[h(\mathcal{X}_{k|k-1})]\}\{h(\mathcal{X}_{k|k-1})-E[h(\mathcal{X}_{k|k-1})]\}^T]+E_{\mathcal{X}_{k|k-1},r'_m} [\{h(\mathcal{X}_{k|k-1})-E[h(\mathcal{X}_{k|k-1})]\}(\mu'-r'_m)^T]\\&+E_{\mathcal{X}_{k|k-1},r'_m} [(\mu'-r'_m)\{h(\mathcal{X}_{k|k-1})-E[h(\mathcal{X}_{k|k-1})]\}^T]+E[(r'_m-\mu')(r'_m-\mu')^T]\\&=\sum_{j=1}^{N_s}w_j[\mathtt{Y}_{j,k|k-1}-E[h(\mathcal{X}_{k|k-1})]][\mathtt{Y}_{j,k|k-1}-E[h(\mathcal{X}_{k|k-1})]]^T+\alpha' R'_{k},
			\end{split}
		\end{equation}
		where $\mathtt{Y}_{j,k|k-1}=h(\mathtt{X}_{j,k|k-1})=h(S_{k|k-1}\xi_j+\hat{\mathcal{X}}_{k|k-1})$, again where $S_{k|k-1}S_{k|k-1}^T=P_{k|k-1}$ and $E[h(\mathcal{X}_{k|k-1})]=\sum_{j=1}^{N_s}w_j\mathtt{Y}_{j,k|k-1}$. So,
		\begin{equation*}
			B_k=(\mathcal{Y}_{k}-E[h(\mathcal{X}_{k|k-1})]-\mu')(\mathcal{Y}_{k}-E[h(\mathcal{X}_{k|k-1})]-\mu')^T+\sum_{j=1}^{N_s}w_j[\mathtt{Y}_{j,k|k-1}-E[h(\mathcal{X}_{k|k-1})]][\mathtt{Y}_{j,k|k-1}-E[h(\mathcal{X}_{k|k-1})]]^T+\alpha' R'_{k}.
		\end{equation*}
	\end{proof} 
	\textbf{Lemma 3:} $D_k$ of \eqref{Eq_DkFirst} can be evaluated as
	\begin{equation}
		D_k=R'_k+\dfrac{1}{\alpha'}(\hat{\mu}'_{k|k-1}-\mu')(\hat{\mu}'_{k|k-1}-\mu')^T,
	\end{equation}
	such that $\hat{\mu}'_{k|k-1}$ is the prior estimate of $\mu$.

	\begin{proof}
		Evaluating $D_k$,
		\begin{equation*}
			\begin{split}
				D_k&=\dfrac{1}{\alpha'}E_{r'_m}[(r'_m-\mu')(r'_m-\mu')^T]\\&
				=\dfrac{1}{\alpha'}E[(r'_m-\hat{\mu}'_{k|k-1}+\hat{\mu}'_{k|k-1}-\mu')(r'_m-\hat{\mu}'_{k|k-1}+\hat{\mu}'_{k|k-1}-\mu')^T]\\&
				=\dfrac{1}{\alpha'}E[(r'_m-\hat{\mu}'_{k|k-1})(r'_m-\hat{\mu}'_{k|k-1})^T+(r'_m-\hat{\mu}'_{k|k-1})(\hat{\mu}'_{k|k-1}-\mu')^T\\&+(\hat{\mu}'_{k|k-1}-\mu')(r'_m-\hat{\mu}'_{k|k-1})^T+(\hat{\mu}'_{k|k-1}-\mu')(\hat{\mu}'_{k|k-1}-\mu')^T]\\&
				=\dfrac{1}{\alpha'}E[(r'_m-\hat{\mu}'_{k|k-1})(r'_m-\hat{\mu}'_{k|k-1})^T+(\hat{\mu}'_{k|k-1}-\mu')(\hat{\mu}'_{k|k-1}-\mu')^T]\\&
				=R'_k+\dfrac{1}{\alpha'}(\hat{\mu}'_{k|k-1}-\mu')(\hat{\mu}'_{k|k-1}-\mu')^T.
			\end{split}
		\end{equation*}
	\end{proof} 
	
	The cross covariance of the state and the measurement is evaluated as in deterministic sample point filters,
	\begin{equation}\label{Eq_Pxy}
		P_{\mathcal{X}\mathcal{Y}}=\sum_{j=1}^{N_s}w_j[\mathtt{X}_{j,k|k-1}-\hat{\mathcal{X}}_{k|k-1}][\mathtt{Y}_{j,k|k-1}-E[h(\mathcal{X}_{k|k-1})]]^T.
	\end{equation}
	The measurement error covariance is evaluated as
	\begin{equation}
		P_{\mathcal{Y}\mathcal{Y}}=\sum_{j=1}^{N_s}w_j[\mathtt{Y}_{j,k|k-1}-E[h(\mathcal{X}_{k|k-1})]][\mathtt{Y}_{j,k|k-1}-E[h(\mathcal{X}_{k|k-1})]]^T+R_k.
	\end{equation}
	Finally, the posterior mean the posterior covariance can be evaluated after calculating the Kalman gain, $K_k=P_{\mathcal{X}\mathcal{Y}}/P_{\mathcal{Y}\mathcal{Y}}$ as follows,
	\begin{equation}\label{PosteriorMeanVB}
		\hat{\mathcal{X}}_{k|k}=\hat{\mathcal{X}}_{k|k-1}+K_k(\mathcal{Y}_k-E[h(\mathcal{X}_{k|k-1})]-\mu),
	\end{equation}
	and
	\begin{equation}
		P_{k|k}=P_{k|k-1}-K_kP_{yy}K_k^T.
	\end{equation}

	\subsection{Fixed point iteration}
	The approximate parameters of the unknown measurement noise statistics are evaluated from their respective guess values using Eqns. \eqref{Eq_R} to \eqref{Eq_U} with the help of the fixed point iteration method which is a numerical method of finding approximate solutions.
	It is denoted by $i$ which represents $i$th iteration as can be seen from the Algorithm \ref{Algorithm}. Initially, when $i=0$, the $\hat{\mathcal{X}}_{k|k}^0=\hat{\mathcal{X}}_{k|k-1}$ and $P_{k|k}^0=P_{k|k-1}$.
	The $R_k$ in \eqref{Eq_R} is updated as
	\begin{equation}\label{Eq_Ri}
		R_k^{i+1}=\dfrac{\hat{U}'_{k|k-1}}{\hat{u}'_{k|k-1}-m-1}.
	\end{equation}
	The $\mu$ of \eqref{Eq_mu} is updates as $\mu^{i+1}$ evaluated as,
	\begin{equation}
		\mu^{i+1}=\dfrac{\mu'+\alpha'(\mathcal{Y}_k-\sum_{j=1}^{N_s}w_j\mathtt{Y}^i_{j,k|k})}{\alpha'+1},
	\end{equation}
	where $\mathtt{Y}^i_{j,k|k}=h(\mathtt{X}^i_{j,k|k})=h(S^i_{k|k}\xi_j+\hat{\mathcal{X}}^i_{k|k})$, where $S^i_{k|k}(S_{k|k}^i)^T=P_{k|k}^i$.
	Then, $B_k^{i+1}$ is updated as, 
	\begin{equation}
		B_k^{i+1}=(\mathcal{Y}_{k}-h(\hat{\mathcal{X}}_{k|k}^i)-\mu^{i+1})(\mathcal{Y}_{k}-h(\hat{\mathcal{X}}_{k|k}^i)-\mu^{i+1})^T+\sum_{j=1}^{N_s}w_j[\mathtt{Y}_{j,k|k}^i-\hat{\mathcal{Y}}^i_{k|k}][\mathtt{Y}_{j,k|k}^i-\hat{\mathcal{Y}}^i_{k|k}]^T+\alpha R^{i+1}_{k},
	\end{equation}
	where $\hat{\mathcal{Y}}^i_{k|k}=\sum_{j=1}^{N_s}w_j\mathtt{Y}^i_{j,k|k}$. $D_k^{i+1}$ is updated as 
	\begin{equation}
		D_k^{i+1}=R^{i+1}_k+\dfrac{1}{\alpha'}({\mu}^{i+1}-\mu')({\mu}^{i+1}-\mu')^T,
	\end{equation}
	Using, $B_k^{i+1}$ and $D_k^{i+1}$, the $\hat{U}_{k|k-1}^{i+1}$ is evaluated as,
	\begin{equation}\label{Eq_Ui}
		\hat{U}_{k|k-1}^{i+1}=\hat{U}'_{k|k-1}+B_k^{i+1}+D_k^{i+1}.
	\end{equation}
	Then the $P_{\mathcal{YY}}$ is updated as $P^{i+1}_{\mathcal{YY}}$ as follows,
	\begin{equation}
		P_{\mathcal{Y}\mathcal{Y}}^{i+1}=\sum_{j=1}^{N_s}w_j[\mathtt{Y}_{j,k|k-1}-E[h(\mathcal{X}_{k|k-1})]][\mathtt{Y}_{j,k|k-1}-E[h(\mathcal{X}_{k|k-1})]]^T+R^{i+1}_k.
	\end{equation}
	So the Kalman gain at each iteration is evaluated as $K_k^{i+1}=P_{\mathcal{XY}}/P_{\mathcal{YY}}^{i+1}$, further using which the posterior mean of the state is evaluated for each iteration,
	\begin{equation}
		\hat{\mathcal{X}}_{k|k}^{i+1}=\hat{\mathcal{X}}_{k|k-1}+K_k^{i+1}(\mathcal{Y}_k-E[h(\mathcal{X}_{k|k-1})]-\mu^{i+1}),
	\end{equation}
	and its error covariance is calculated as,
	\begin{equation}
		P_{k|k}^{i+1}=P_{k|k-1}-K_k^{i+1}P_{yy}{K_k^{i+1}}^T.
	\end{equation}
	When the difference, $\hat{\mathcal{X}}^{i+1}_{k|k}-\hat{\mathcal{X}}_{k|k}^i$ is less than $\zeta$ (small value chosen by practitioners), the loop for fixed point iteration terminates denoting not much significant change in the posterior mean occurs in the next iterations. Such a state is generally achieved quickly within a few number of iterations (say, 6 to 7). In our work, we chose the value of $\zeta$ to be $10^{-3}$. The posterior mean at the final iteration say $\hat{\mathcal{X}}_{k|k}^{N}$ and its error covariance, $P_{k|k}^{N}$ is considered as the estimated mean and the error covariance \emph{i.e.,}  $\hat{\mathcal{X}}_{k|k}=\hat{\mathcal{X}}_{k|k}^{N}$ and $P_{k|k}=P_{k|k}^{N}$. Further, the confidence parameter $\alpha'$ and the degree of freedom $\hat{u}'_{0|0}$ is tuned maximizing the likelihood as shown in Algorithm \ref{OptiAlgorithm}.
	
	
	\begin{algorithm}[h!]
		\caption{VB approach for unknown noise covariance}\label{Algorithm}
		\begin{algorithmic}[1]
			\STATE Inputs: $\hat{\mathcal{X}_{0|0}}, P_{0|0}, r'_m, \alpha', \hat{u}'_{0|0}, \hat{U}'_{0|0}=(\hat{u}'_{0|0}-m-1)R'_0$
			\FOR{k=$1$ to $TT$}
			\STATE Time Update:\\
			$\hat{\mathcal{X}}_{k|k-1}=F\hat{\mathcal{X}}_{k-1|k-1}-\mho_{k-1,k}$\\  
			$P_{k|k-1}=FP_{k-1|k-1}+Q_k$
			\STATE Measurement Update:
			\STATE Calculate: $S_{k|k-1}=chol(P_{k|k-1})$.
			\STATE Evaluate sigma points: $\mathtt{X}_{j,k|k-1}=S_{k|k-1}\xi_j+\hat{\mathcal{X}}_{k|k-1}$, where $j=1,2,...,N_s$.
			\STATE Predicted measurement at each point, ${\mathtt{Y}}_{k|k-1}=h(\mathtt{X}_{j,k|k-1})$.
			\STATE Predicted measurement estimation: $E[h(\mathcal{X}_{k|k-1})]=\sum_{j=1}^{N_s}w_j\mathtt{Y}_{j,k|k-1}$.
			\STATE Calculate measurement and state error cross covariance: $P_{\mathcal{X}\mathcal{Y}}=\sum_{j=1}^{N_s}w_j[\mathtt{X}_{j,k|k-1}-\hat{\mathcal{X}}_{k|k-1}][\mathtt{Y}_{j,k|k-1}-E[h(\mathcal{X}_{k|k-1})]]^T$.
			\STATE Fixed point iteration:
			Initialise $\hat{u}'_{k|k-1}=\hat{u}'_{k-1|k-1}$, $\hat{U}'_{k|k-1}=\hat{U}'_{k-1|k-1}$, $N=0$, $i=0$.
			\WHILE{N=0}
			\STATE Evaluate: $R^{i+1}_{k}$ from \eqref{Eq_R}.
			\STATE Evaluate: $\mu^{i+1}$ and $\alpha^{i+1}$ from \eqref{Eq_mu} and \eqref{Eq_alpha}, respectively.
			\STATE Evaluate: $\hat{u}^{i+1}_{k|k-1}$ from Eqn. \eqref{Eq_u}, $\hat{U}^{i+1}_{k|k-1}$ from Eqn. \eqref{Eq_U} using $B_k^i$ from Eqn. \eqref{Eq_Bk}.\\
			\STATE Calculate measurement error covariance: $P_{\mathcal{Y}\mathcal{Y}}^{i+1}=\sum_{j=1}^{N_s}w_j[\mathtt{Y}_{j,k|k-1}-E[h(\mathcal{X}_{k|k-1})]][\mathtt{Y}_{j,k|k-1}-E[h(\mathcal{X}_{k|k-1})]]^T+R_k^{i+1}$.
			
			\STATE Kalman gain: $K_k^{i+1}=P_{\mathcal{X}\mathcal{Y}}/P_{\mathcal{Y}\mathcal{Y}}^{i+1}$.
			\STATE Posterior mean at $(i+1)$th step: $\hat{\mathcal{X}}_{k|k}^{i+1}=\hat{\mathcal{X}}_{k|k-1}+K_k^{i+1}(\mathcal{Y}_k-E[h(\mathcal{X}_{k|k-1})]-\mu^{i+1})$
			\STATE Posterior error covariance at $(i+1)$th step: $P^{i+1}_{k|k}=P_{k|k-1}-K_k^{i+1}P^{i+1}_{yy}(K_k^{i+1})^T$.
			\IF{$\hat{\mathcal{X}}_{k|k}^{i+1}-\hat{\mathcal{X}}_{k|k}^{i}<10^{-3}$}
			\STATE $N=i+1$.
			\ELSE \STATE $i=i+1$.
			\ENDIF
			\ENDWHILE
			\STATE $\hat{\mathcal{X}}_{k|k}=\hat{\mathcal{X}}_{k|k}^{N}, P_{k|k}=P_{k|k}^{N}, \hat{u}'_{k|k}=\hat{u}_{k|k-1}^N,\hat{U}'_{k|k}=\hat{U}_{k|k-1}^N$.
			\ENDFOR
		\end{algorithmic}
	\end{algorithm}
	
	\begin{algorithm}[h!]
		\caption{VB approach for unknown noise covariance with optimized tuning parameters}\label{OptiAlgorithm}
		\begin{algorithmic}[1]
			\STATE Inputs: $\hat{\mathcal{X}}_{0|0},P_{0|0},r'_m$.
			\STATE Follow steps 2 to 10 of Algorithm \ref{Algorithm}.
			\WHILE{N=0}
			\IF{$t=1$} 
			\FOR{$\hat{u}'_{0|0}=3$ to $23$}
			\STATE Assume, $\alpha'=1$
			\STATE Evaluate $\hat{U}'_{0|0}=(\hat{u}'_{0|0}-m-1)R'_0$
			\STATE Evaluate steps 11 to 18 of Algorithm \ref{Algorithm}.
			\STATE Evaluate likelihood: $L_{\hat{u}'_{0|0}}=\dfrac{1}{\sqrt{2\pi P_{\mathcal{Y}\mathcal{Y}}^{(i)}}} \exp(-\dfrac{1}{2}\dfrac{(\mathcal{Y}_k-\hat{\mathcal{Y}}_{k|k-1}^{(i)})^2}{P_{\mathcal{Y}\mathcal{Y}}^{(i)}})$
			\ENDFOR
			\STATE $\hat{u}'_{0|0}=\text{arg}\max\limits_{\substack{\hat{u}'_{0|0}}}L_{\hat{u}'_{0|0}}$
			\ENDIF
			\FOR{$\alpha'=1$ to $20$}
			\STATE Evaluate steps 11 to 18 of Algorithm \ref{Algorithm}.
			\STATE Evaluate likelihood: $L_{\alpha'}=\dfrac{1}{\sqrt{2\pi P_{\mathcal{Y}\mathcal{Y}}^{(i)}}} \exp(-\dfrac{1}{2}\dfrac{(\mathcal{Y}_k-\hat{\mathcal{Y}}_{k|k-1}^{(i)})^2}{P_{\mathcal{Y}\mathcal{Y}}^{(i)}})$
			\ENDFOR
			\STATE $\alpha'=\text{arg}\max\limits_{\substack{\alpha'}}L_{\alpha'}$
			
			\STATE Evaluate steps 11 to 23 of Algorithm \ref{Algorithm}.
			\ENDWHILE
			\STATE Repeat steps 25 and 30 of Algorithm \ref{Algorithm}.
		\end{algorithmic}
	\end{algorithm}
	\section{The MAPMLE Technique of Adaptive Filtering}
	The  maximum a posteriori (MAP) technique of estimating both measurement noise mean and covariance does not confirm the positive definiteness of the covariance matrix. So to estimate the measurement noise covariance the maximum likelihood (MLE) of residual based adaptive Kalman filter technique has been implemented that ensures the positive definiteness of the measurement covariance and the measurement noise mean estimation the MAP based technique is considered. Initially the estimation technique is performed with a guess value of the measurement noise mean and covariance at $k=0$. After each measurement update of any traditional filter, the measurement noise mean and the covariance are estimated. The estimated mean and the covariance is used up in the next time and measurement update of the filter.
	\subsection{Measurement noise mean estimation}
	The measurement noise mean can be obtained after each measurement update using the following equation  \cite{zhao2009adaptive}. 
	\begin{equation}\label{rmeanest}
		\hat{r}'_{{m}_k}=\dfrac{1}{k}\sum_{i=1}^{k}(\mathcal{Y}_k-E[h(\mathcal{X}_{k|k-1})])
	\end{equation}
	This is well known as the maximum a posteiori (MAP) technique of estimating the mean of measurement noise.
	\subsection{Measurement noise covariance estimation}
	The adaptive filter based on residual sequence is called residual based R-adaptive method and the one based on innovation sequence is named as innovation based R-adaptive method. The innovation or the residual sequence is computed with a sliding window whose length is considered to be $L$.
	But in the innovation based method the positive definiteness of measurement covariance is not confirmed as it is obtained by the difference of two positive definite matrices as in \cite{mehra1970identification, mohamed1999adaptive, yang2003adaptive} which says \begin{equation}\label{innovation}
		\hat{R}_k=\hat{C}_v-H_kP_{k|k-1}H_K^T,
	\end{equation} where \begin{equation}\label{cv}
		\hat{C}_v=\dfrac{1}{L}\sum_{i=1}^L(\mathcal{Y}_k-\hat{\mathcal{Y}}_k)(\mathcal{Y}_k-\hat{\mathcal{Y}}_k)^T
	\end{equation}
	So instead of using Equation \ref{innovation} another method to estimate measurement noise covariance was proposed in \cite{wang1999stochastic} to guarantee the positive definiteness of $R_k$, which is referred to as the residual method. The $R_k$ is estimated as follows:
	\begin{equation}\label{residual}
		\hat{R}_k=\hat{C}_v+H_kP_{k|k}H_K^T.
	\end{equation}
	In our work, we use the residual method of $R_k$ adaptation.

	
	\section{Properties}
	\subsection{Biasness}
	An estimation technique is said to be unbiased if the expectation of the difference in the estimated state and the true state is zero \emph{i.e.,} if $E[\hat{\mathcal{X}}_{k|k}-\mathcal{X}_k]=0$.
	\subsubsection{Biasness of state for variational Bayesian technique of adaptive filtering}
	The difference in the estimated state and the true state using \eqref{PosteriorMeanVB} of the VB approach can be evaluated as, 
	\begin{equation}
		\begin{split}
			\hat{\mathcal{X}}_{k|k}-\mathcal{X}_k&=\hat{\mathcal{X}}_{k|k-1}+K_k(\mathcal{Y}_k-E[h(\mathcal{X}_{k|k-1})]-\mu)-(F\mathcal{X}_{k-1}+\omega_{k-1}-\mho_{k-1,k})\\&
			=F\hat{\mathcal{X}}_{k-1|k-1}-\mho_{k-1,k}+K_k(\mathcal{Y}_k-E[h(\mathcal{X}_{k|k-1})]-\mu)-(F\mathcal{X}_{k-1}+\omega_{k-1}-\mho_{k-1,k})\\&
			=F(\hat{\mathcal{X}}_{k-1|k-1}-\mathcal{X}_{k-1})-\omega_{k-1}+K_k(\mathcal{Y}_k-E[h(\mathcal{X}_{k|k-1})]-\mu)\\&
		\end{split}
	\end{equation}
	Taking expectations on both sides and using \eqref{Eq_mu} we get,
	\begin{equation}
		\begin{split}\label{Eq_BiasStateNIW}
			E[\hat{\mathcal{X}}_{k|k}-\mathcal{X}_k]&=F(\hat{\mathcal{X}}_{k-1|k-1}-E[\mathcal{X}_{k-1}])-E[\omega_{k-1}]+K_k(E[\mathcal{Y}_k]-E[h(\mathcal{X}_{k|k-1})]-E[\mu])\\&
			=K_k(E[h(\mathcal{X}_{k|k-1})]+r_m-E[h(\mathcal{X}_{k|k-1})]-E[\dfrac{\mu'+\alpha'(\mathcal{Y}_k-E[h(\mathcal{X}_{k|k-1})])}{\alpha'+1}])\\&
			=K_k(r_m-\dfrac{\mu'+\alpha'(E[\mathcal{Y}_k]-E[h(\mathcal{X}_{k|k-1})])}{\alpha'+1})\\&
			=K_k(r_m-\dfrac{\mu'+\alpha'(r_m)}{\alpha'+1})\\&
			=K_k\dfrac{\alpha'r_m+r_m-\mu'-\alpha'r_m}{\alpha'+1}\\&
			=K_k\dfrac{r_m-\mu'}{\alpha'+1}
		\end{split}
	\end{equation}
	As the approximate mean, $\mu'$ approaches the true mean, $r_m$ using fixed point iteration  \eqref{Eq_BiasStateNIW} equates to zero.
	
	\subsubsection{Biasness of state for the MAPMLE technique of adaptive filtering}
	The difference between the estimated state and the true state can be evaluated as,
	\begin{equation}
		\begin{split}
			\hat{\mathcal{X}}_{k|k}-\mathcal{X}_k&=\hat{\mathcal{X}}_{k|k-1}+K_k(\mathcal{Y}_k-E[h(\mathcal{X}_{k|k-1})]-\hat{r}'_{m_k})-\mathcal{X}_{k}\\&
			=F\hat{\mathcal{X}}_{k-1|k-1}-\mho_{k-1,k}+K_k(\mathcal{Y}_k-E[h(\mathcal{X}_{k|k-1})]-\hat{r}'_{m_k})-F\mathcal{X}_{k-1}-\omega_{k-1}+\mho_{k-1,k}\\&
			=F(\hat{\mathcal{X}}_{k-1|k-1}-\mathcal{X}_{k-1})-\omega_{k-1}+K_k(\mathcal{Y}_k-E[h(\mathcal{X}_{k|k-1})]-\hat{r}'_{m_k})
		\end{split}
	\end{equation}
	Taking expectation on both side
	\begin{equation}
		\begin{split}
			E[\hat{\mathcal{X}}_{k|k}-\mathcal{X}_k]&=F(\hat{\mathcal{X}}_{k-1|k-1}-E[\mathcal{X}_{k-1}])-E[\omega_{k-1}]+K_k(E[\mathcal{Y}_k]-E[h(\mathcal{X}_{k|k-1})]-\hat{r}'_{m_k})\\&
			=K_k(E[h(\mathcal{X}_{k|k-1})] + r_m -E[h(\mathcal{X}_{k|k-1})]-\hat{r}'_{m_k})\\&
			=K_k(r_m-\hat{r}'_{m_k})
		\end{split}
	\end{equation}
	As the approximate estimate of $\hat{r}'_{m_k}$ approaches $r_m$, the biasness of the state equates to zero.


	\section{Simulation Results}
\begin{figure}
		\begin{subfigure}[b]{0.5\textwidth}
			\centering
			\includegraphics[width=8cm,height=6.5cm]{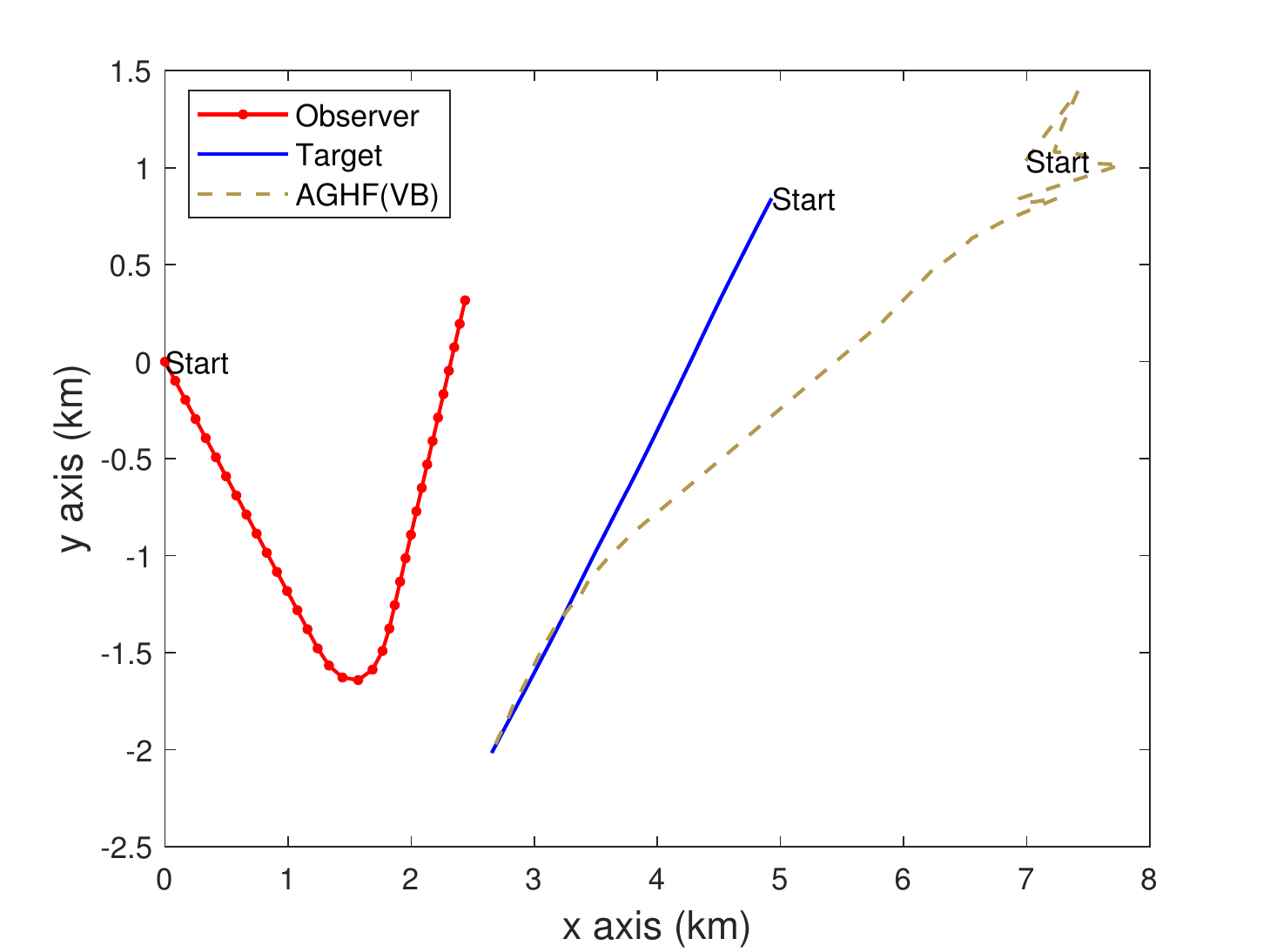}
			\caption{}
			\label{fig_scenario1}
		\end{subfigure}		
		\begin{subfigure}[b]{0.5\textwidth}
			\centering
			\includegraphics[width=8cm,height=6.5cm]{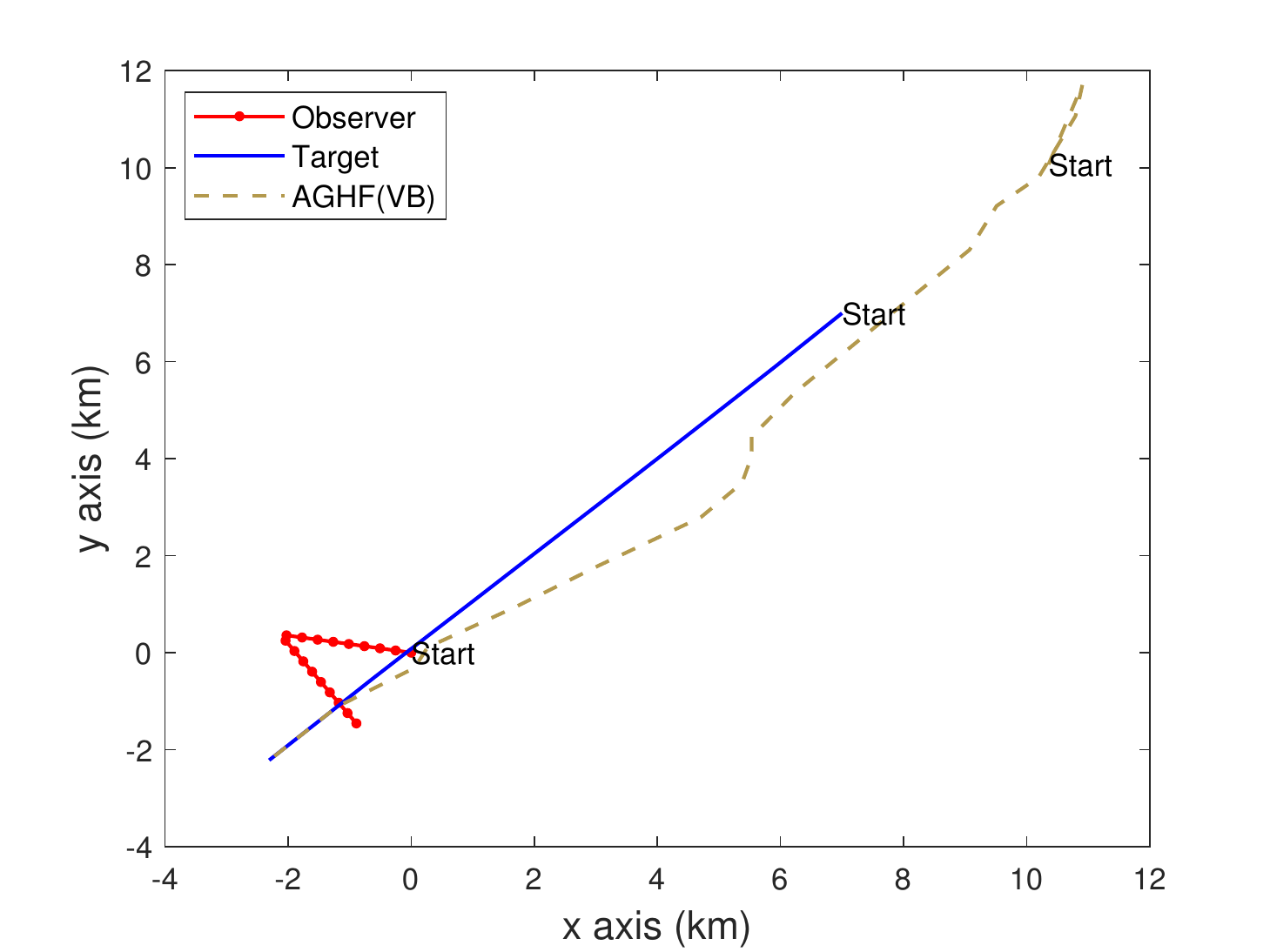}
			\caption{}
			\label{fig_scenario2}			
		\end{subfigure}
		\caption{Tracking scenario for (a) moderately nonlinear scenario (Scenario I), and  (b) highly nonlinear scenario (Scenario II)}
	\end{figure}

	Here two scenarios, one moderately nonlinear and one highly nonlinear, are considered as shown in Figure \ref{fig_scenario1} and \ref{fig_scenario2} respectively. In both scenarios, the target moves in a nearly straight line motion with constant velocity. The ownship maneuvers from $13$th min to $17$th min for the moderately nonlinear scenario and at $15$th min for the highly nonlinear scenario to make the system observable. While maneuvering, the rate of change of bearing for the highly nonlinear scenario is high thus making it more difficult for the sub-optimal filters to track. The parameters of the scenarios are given in Table \ref{parameters}, where Scenario I denotes the moderately nonlinear scenario and Scenario II specifies the highly nonlinear scenario parameters. 
	The sampling time is considered to be $5$ s \emph{i.e.} $\Delta=5$ s and the observation is done for a total time period of 30 min.  The process noise intensity, $\bar{q}$ is considered to be $1.944 \times 10^{-6}$ km$^2$/min$^3$ for both scenarios. The standard deviation (S.D.) of initial range, $\sigma_r$, initial target speed, $\sigma_s$, initial bearing, and initial course angle, $\sigma_c$ which are considered in this paper are also given in Table \ref{parameters}. 
	
	Two cases are considered for each scenario based on the stationary and varying measurement noise covariance. Case I, consists of stationary $\sigma_{\theta_k}$ which is $1.5^o\forall k$ for Scenario I and $2^o\forall k$ for Scenario II and Case II has varying $\sigma_{\theta_k}$ such that it varies linearly with the distance of the target from the ownship \emph{i.e.,} more the distance, more is the value of $\sigma_{\theta_k}$ for both the scenarios. The maximum and the minimum values for $\sigma_{\theta_k}$ in Case II are considered to be $4^o$ and $1.5^o$, respectively. We can evaluate ${\sigma_{\theta_k}}$ at the $k$th time step based on the distance between the target and the ownship for Case II using a straight line equation, ${\sigma_{\theta_k}}=sd_k+c$, where $s$ represents slope, $d_k$ is the distance of the target from the ownship at the $k$th time instant and $c$ is the $y$ axis intercept. For maximum value of $d_k$ say, ${d_k}_{max}$, ${\sigma_{\theta_k}}=4^o$ and for minimum value of $d_k$
	say, ${d_k}_{min}$, ${\sigma_{\theta_k}}=1.5^o$. Putting these values in the straight line equation and solving them we get $s=\dfrac{4^o-1.5^o}{{d_k}_{max}-{d_k}_{min}}$ and $c=-\dfrac{4^o{d_k}_{min}-1.5^o{d_k}_{max}}{{d_k}_{max}-{d_k}_{min}}$. Thus, assuming the values of ${d_k}_{max}$ and ${d_k}_{min}$ are known for both the scenarios, we can say,
	\begin{equation}
		{\sigma_{\theta_k}}=\dfrac{(4^o-1.5^o)d_k-4^o{d_k}_{min}+1.5{d_k}_{max}}{{d_k}_{max}-{d_k}_{min}}.
	\end{equation}
	The value for true measurement mean is considered to be $0.1^o$ for both scenarios in both cases. The initial guess value of the measurement noise mean and the measurement noise covariance used for adaptive filtering is considered to be half of their true initial values. 
	
	\begin{table}
		\renewcommand{\arraystretch}{1.4}
		\setlength{\tabcolsep}{5pt}
		\centering 
		\caption{Parameters of the scenarios}
		\begin{tabular}{|c|c|c|}
			\hline
			Parameters   & Scenario I &  Scenario II\\ \hline
			Initial range ($r$)& 5 km & 10 km 	\\ \hline
			Target speed ($s$)& 4 knots & 15 knots\\ \hline
			Target course & $-140^o$ & $-135.4^o$ \\ \hline		
			Observer speed & 5 knots & 5 knots\\ \hline
			Observer initial course & $140^o$ & $-80^o$ \\ \hline		
			Observer final course & $20^o$ & $146^o$ \\ \hline	
			Observer maneuver & From $13^{th}$ to $17^{th} min$ & $15^{th}$ min\\ \hline
			Initial range S.D. $(\sigma_r)$ & 2 km & 4 km \\ \hline
			Initial target speed S.D. $(\sigma_s)$ & 2 knots & 2 knots\\ \hline
			Initial course S.D. $(\sigma_{c})$ & $\pi/\sqrt{12}$ & $\pi/\sqrt{12}$\\ \hline
			Measurement noise mean $(r_m)$ & $0.1^o$ & $0.1^o$ \\ \hline
			
		\end{tabular}
		\label{parameters}
	\end{table}
	\subsection{Performance matrices}
	The performance of the filters is analyzed using the following performance matrices.
	\subsubsection{Root mean square error}
	The root mean square error (RMSE) of position and velocity  at the $k$-th time instant is evaluated as
	\begin{equation}
		RMSE_{pos,k}=\sqrt{\dfrac{1}{M}\Sigma_{l=1}^M(x_k^{t,l}-\hat{x}_k^{t,l})^2+(y_k^{t,l}-\hat{y}_k^{t,l})^2}
	\end{equation}
	and 
	\begin{equation}
		RMSE_{vel,k}=\sqrt{\dfrac{1}{M}\Sigma_{l=1}^M(\dot{x}_k^{t,l}-\hat{\dot{x}}_k^{t,l})^2+(\dot{y}_k^{t,l}-\hat{\dot{y}}_k^{t,l})^2}
	\end{equation} respectively. In this paper, we considered 500 Monte Carlo runs to evaluate the RMSE.
	\subsubsection{Track loss percentage}
	The percentage of track loss is obtained from the fact that any estimate whose terminal error in position is beyond 0.2km is considered as diverged track and hence eliminated. Table \ref{tracklossCaseI} and \ref{tracklossCaseII} show the track loss percentage using all the filters for both scenarios for 10,000 Monte Carlo runs.
	\subsubsection{Bias norm}
	The bias norm at the $k$th time instant of a filter is evaluated as
	\begin{equation}
		\text{Bias norm}_k=\Bigg|\Bigg|\dfrac{1}{M}\sum_{l=1}^{M}\hat{\mathcal{X}}^l_{k|k}-\dfrac{1}{M}\sum_{l=1}^{M}\mathcal{X}^l_k\Bigg|\Bigg|_2,
	\end{equation}
	where $||\cdot||_2$ represents the vector norm, $M$ is the total number of Monte Carlo runs. Here we have considered 500 Monte Carlo runs to evaluate the bias norm. 
	\subsubsection{Average normalized estimation error squared (ANEES)}
	The normalized estimation error squared at the $k$th instant of time, $\text{NEES}_k$ is evaluated as
	\begin{equation}
		\text{NEES}_k=(\mathcal{X}_k-\hat{\mathcal{X}}_{k|k})^TP_{k|k}^{-1}(\mathcal{X}_k-\hat{\mathcal{X}}_{k|k}).
	\end{equation}
	If the estimation technique is consistent, the $\text{NEES}$ should follow $\chi^2$-distribution with $n$ degree of freedom and a mean of $n$ \emph{i.e.,} $E[\text{NEES}]=n$, where $n$ is the state dimension. The average of NEES is evaluated as
	\begin{equation}
		\text{ANEES}_k=\dfrac{1}{nM}\sum_{l=1}^{M}\text{NEES}_k^l.
	\end{equation}
	If ANEES$_k\in[b_1,b_2]$, where $b_1$ and $b_2$ represent the lower and upper bound of the 95\% probability region, the filter is said to be consistent. If ANEES$_k>b_2$ the filter is said to be optimistic, as the posterior covariance, $P_{k|k}$ is too small, and if ANEES$_k<b_1$ the filter is said to be pessimistic. The values of $b_1$ and $b_2$ are evaluated as in \cite{bar2004estimation}. Here we have considered 500 Monte Carlo runs to evaluate the ANEES.
	\subsection{Performance of filters}
	\subsubsection{Case I: Static measurement noise covariance}
\begin{figure}
		\begin{subfigure}[b]{0.5\textwidth}
			\centering
			\includegraphics[width=8cm,height=6.5cm]{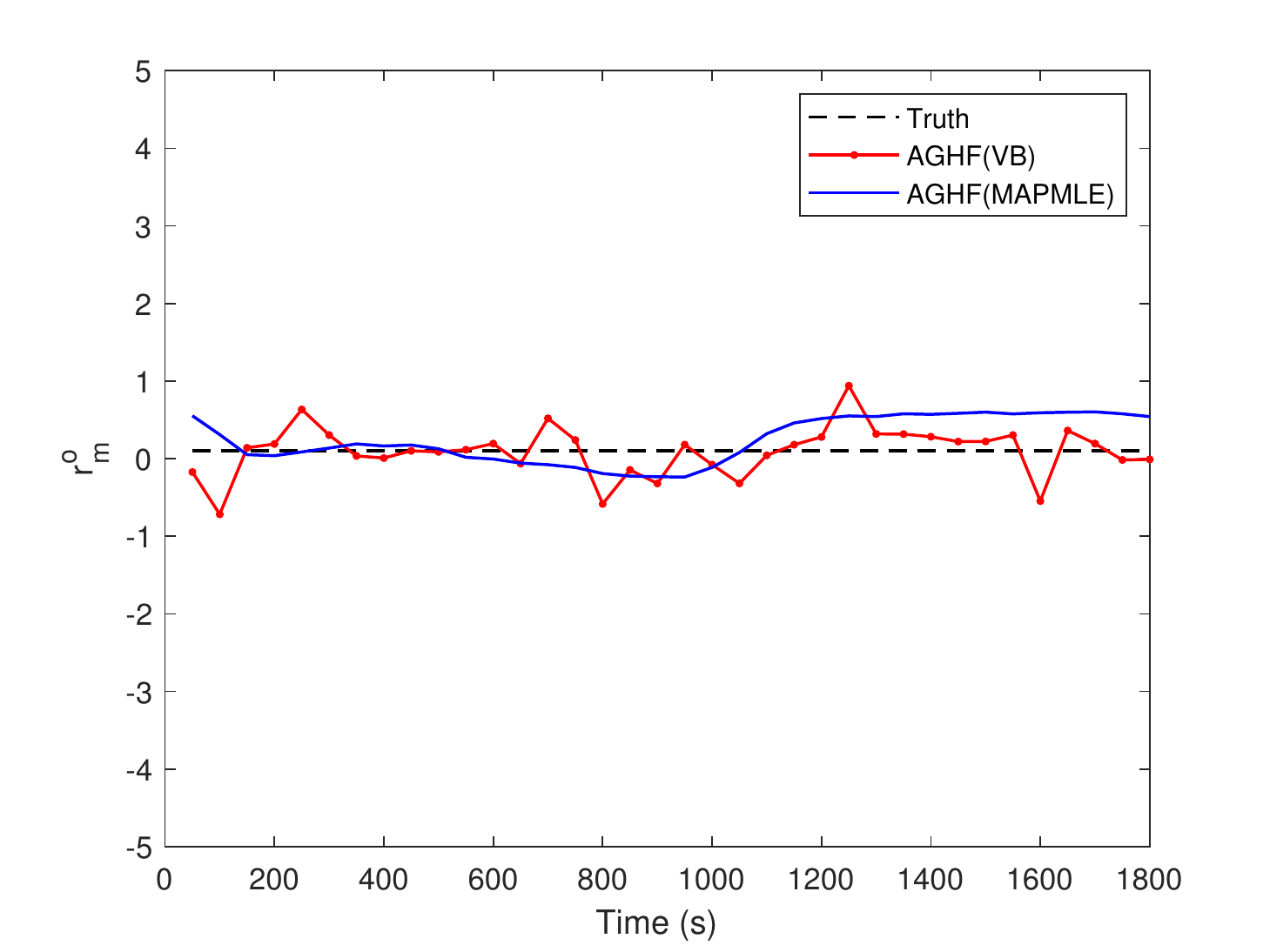}
			\caption{}
			\label{fig_rmStatRSc1}
		\end{subfigure}		
		\begin{subfigure}[b]{0.5\textwidth}
			\centering
			\includegraphics[width=8cm,height=6.5cm]{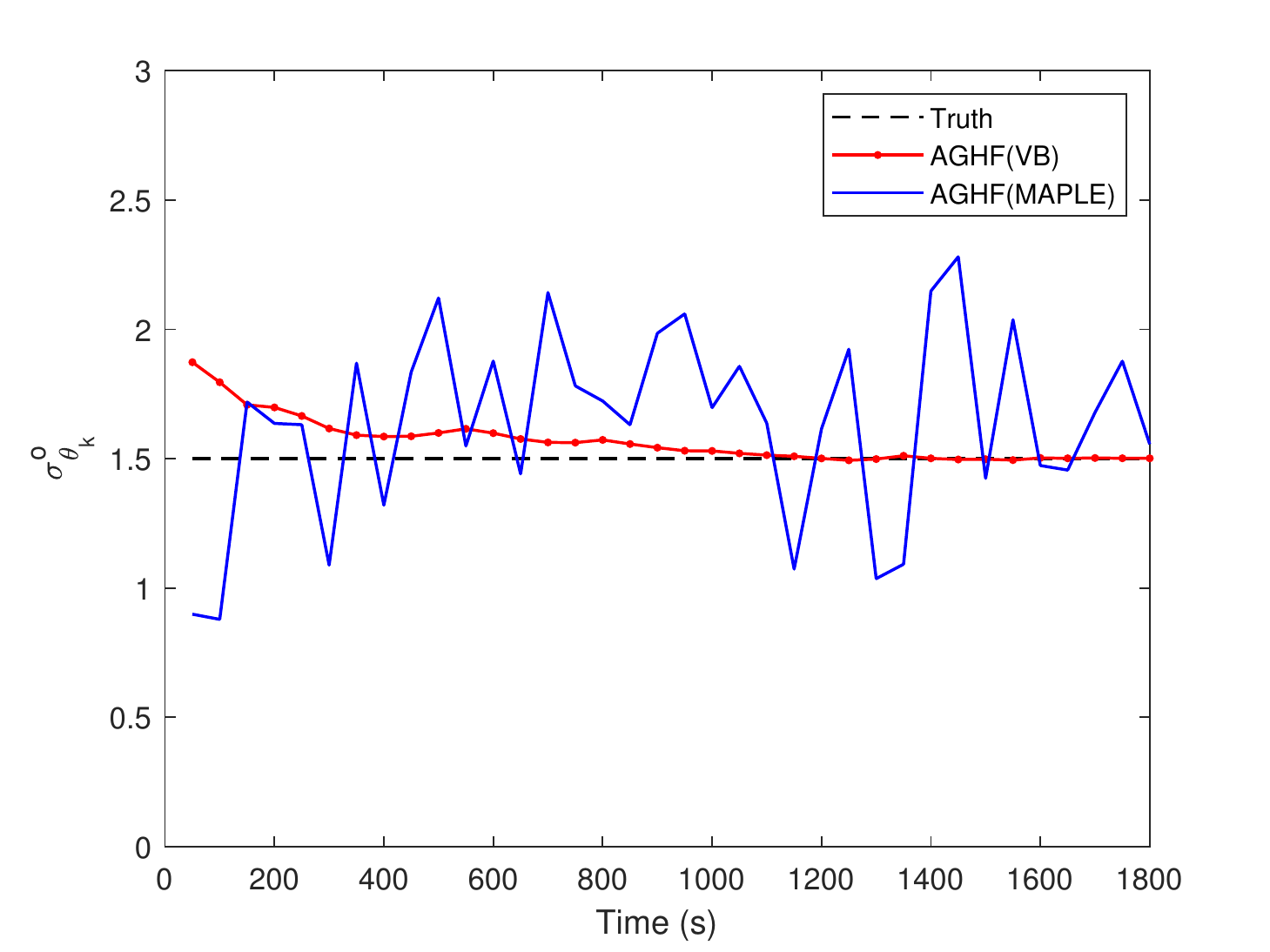}
			\caption{}
			\label{fig_SigThStatRSc1}			
		\end{subfigure}
		\caption{Estimated (a)$r_m$ and (b) $\sigma_{\theta_k}$ for Scenario I, Case I.}
	\end{figure}

\begin{figure}
		\begin{subfigure}[b]{0.5\textwidth}
			\centering
			\includegraphics[width=8cm,height=6.5cm]{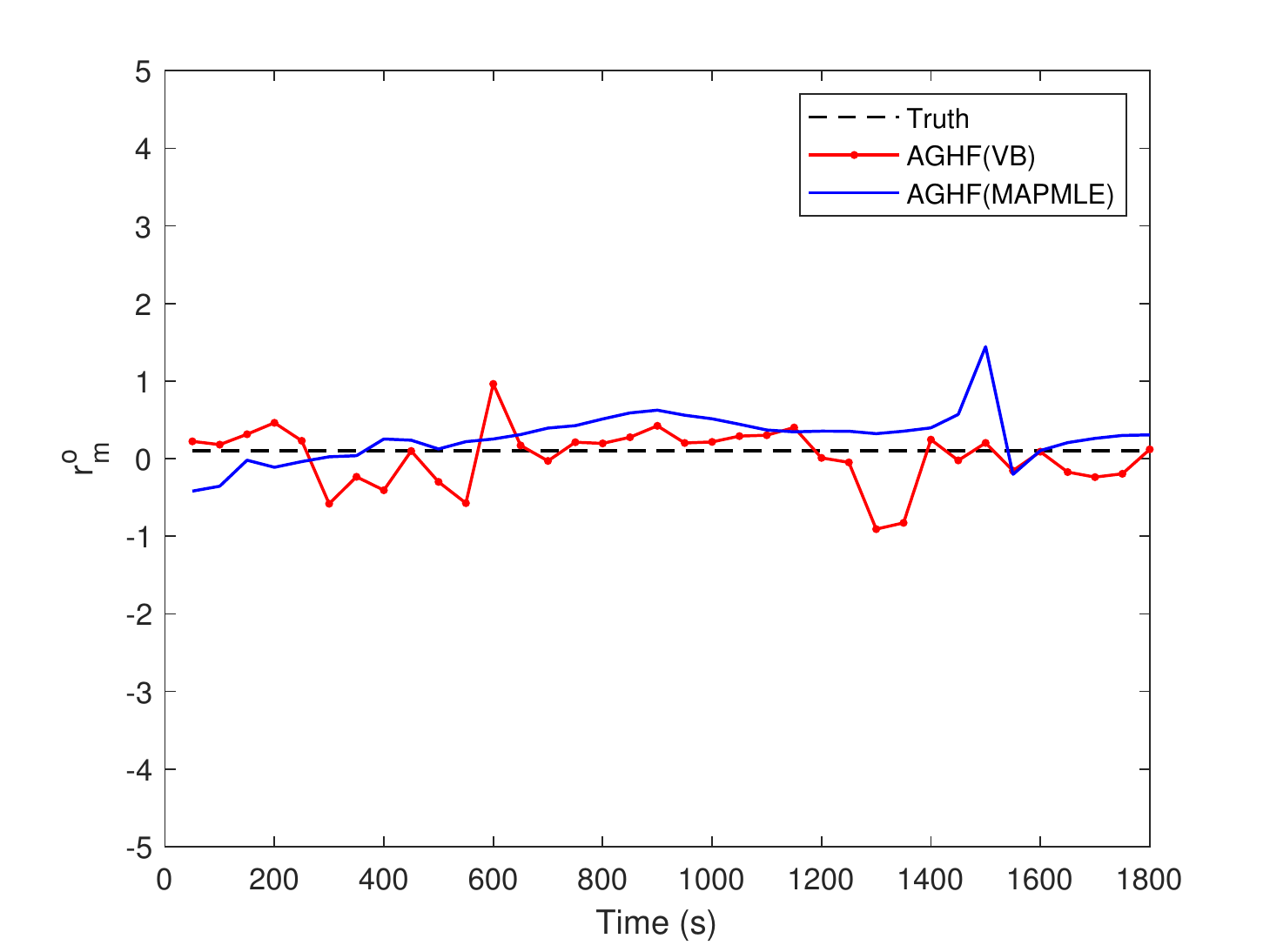}
			\caption{}
			\label{fig_rmStatRSc2}
		\end{subfigure}		
		\begin{subfigure}[b]{0.5\textwidth}
			\centering
			\includegraphics[width=8cm,height=6.5cm]{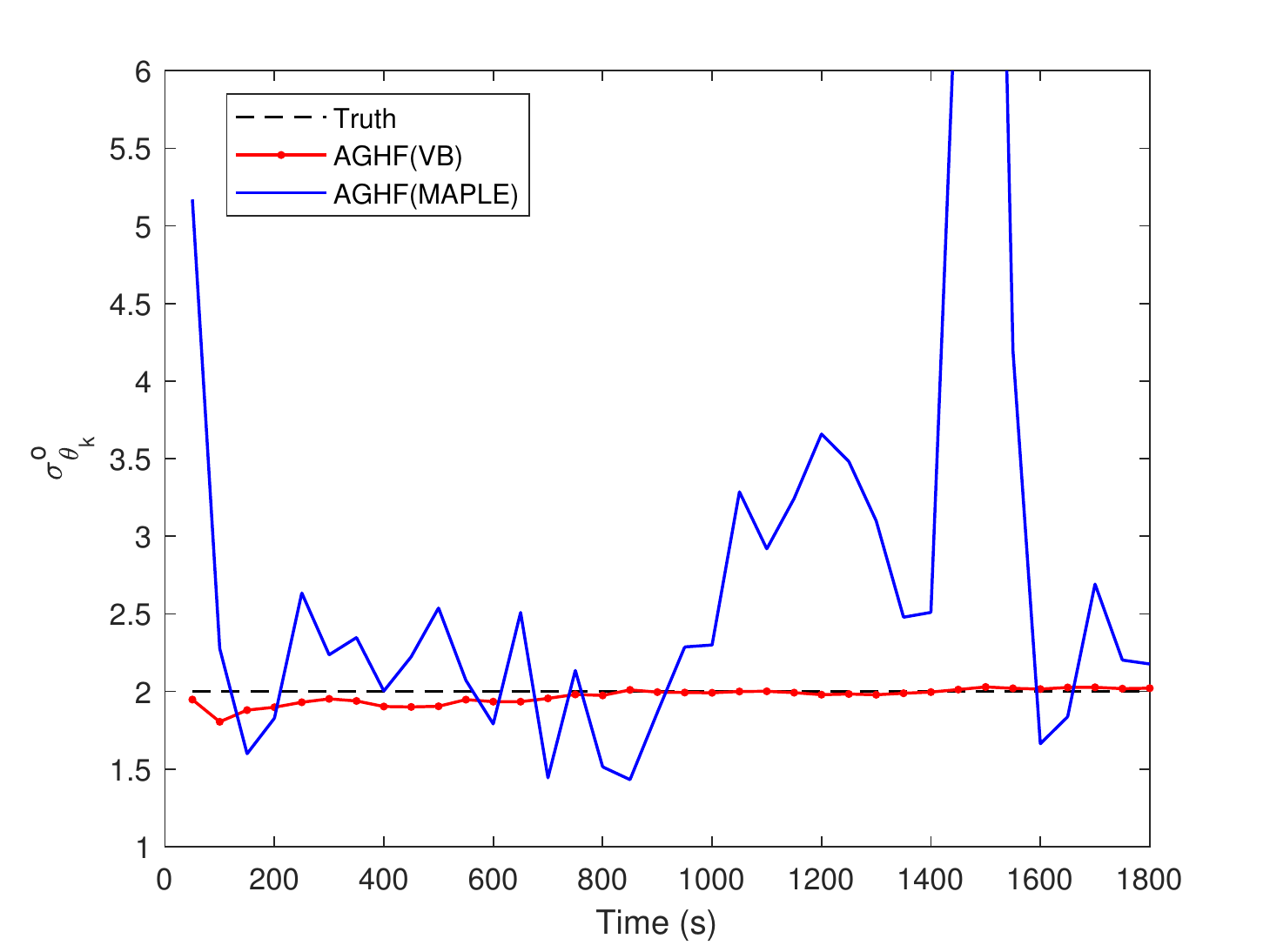}
			\caption{}
			\label{fig_SigThStatRSc2}			
		\end{subfigure}
		\caption{Estimated (a)$r_m$ and (b) $\sigma_{\theta_k}$ for Scenario II, Case I.}
	\end{figure}

	The single run estimation plot for measurement noise mean and standard deviation are shown in \ref{fig_rmStatRSc1} and \ref{fig_SigThStatRSc1}, respectively for Scenario I and \ref{fig_rmStatRSc2} and \ref{fig_SigThStatRSc2}, respectively for Scenario II. The black dashed line marks the true value of the measurement noise mean and standard deviation, the blue line denoted the AGHF following the MAPMLE technique and the red line denoted the AGHF following the VB technique. These figures show that the truth is followed by both the adaptive filters while estimating both the measurement noise mean and the standard deviation. 
	
\begin{figure}
		\begin{subfigure}[b]{0.5\textwidth}
			\centering
			\includegraphics[width=8cm,height=6.5cm]{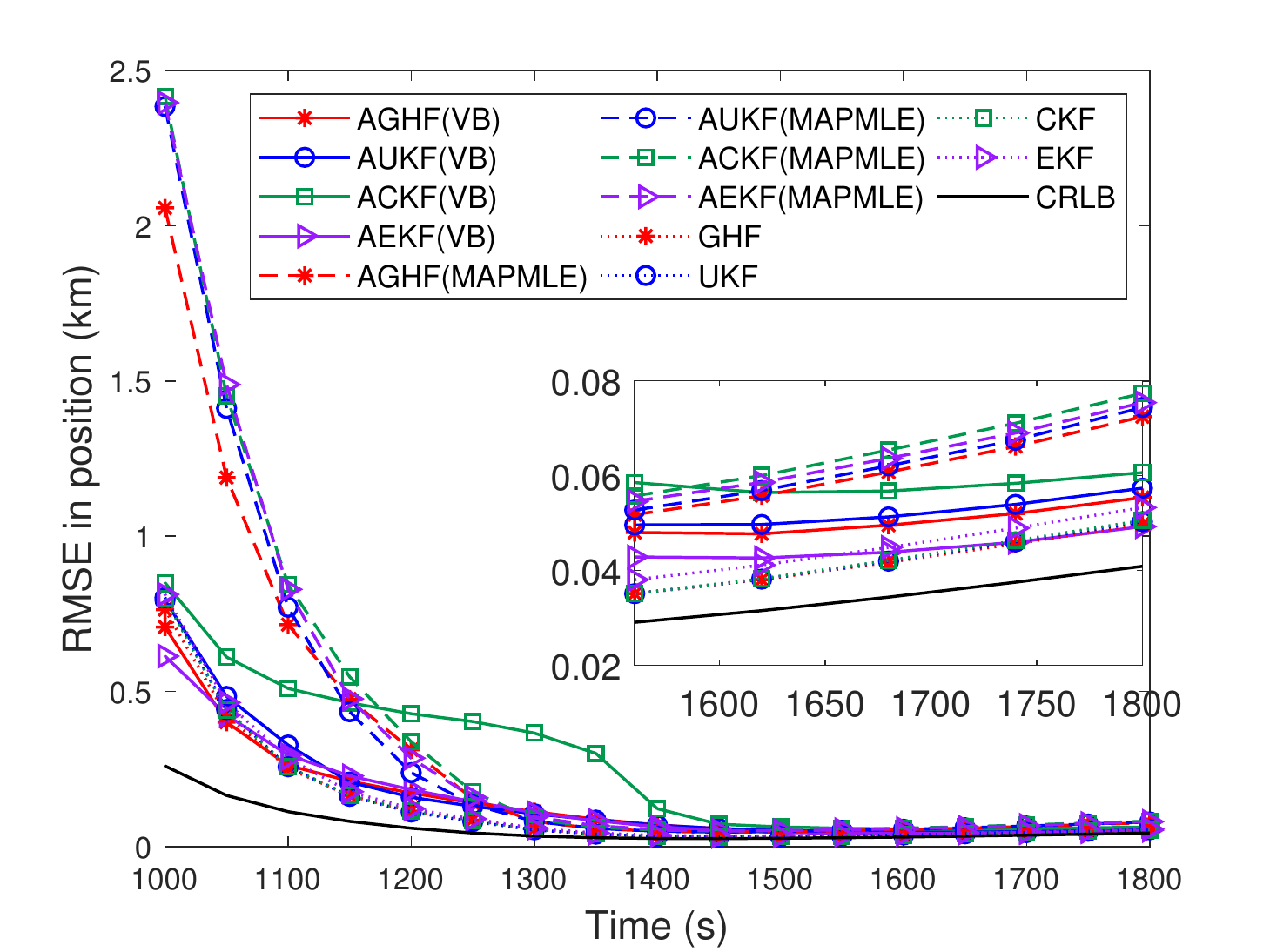}
			\caption{}
			\label{fig_RMSEpstatRSc1}
		\end{subfigure}		
		\begin{subfigure}[b]{0.5\textwidth}
			\centering
			\includegraphics[width=8cm,height=6.5cm]{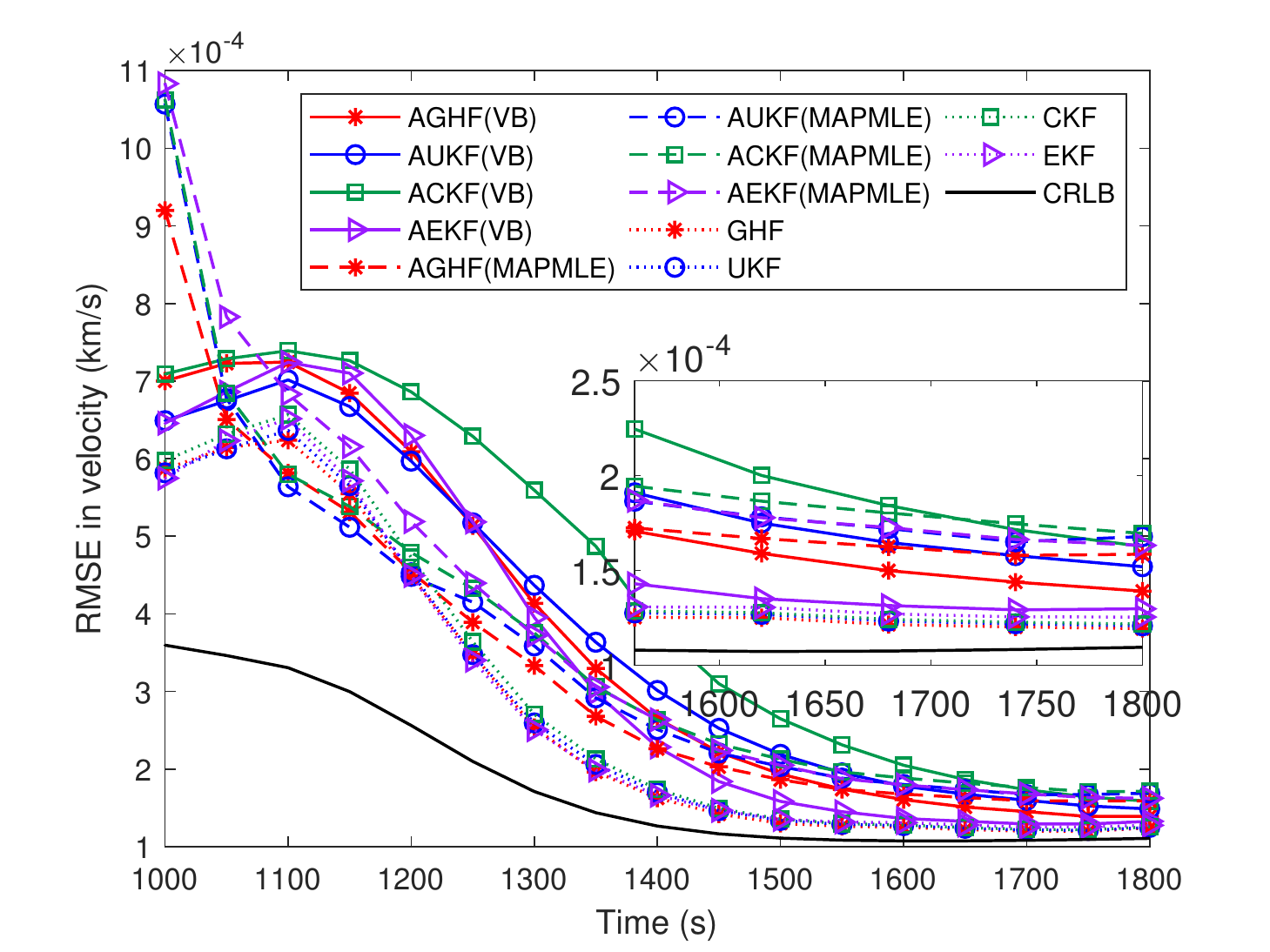}
			\caption{}
			\label{fig_RMSEvstatRSc1}			
		\end{subfigure}
		\caption{RMSE in (a) position, and  (b) velocity for Scenario I, Case I.}
	\end{figure}

\begin{figure}
		\begin{subfigure}[b]{0.5\textwidth}
			\centering
			\includegraphics[width=8cm,height=6.5cm]{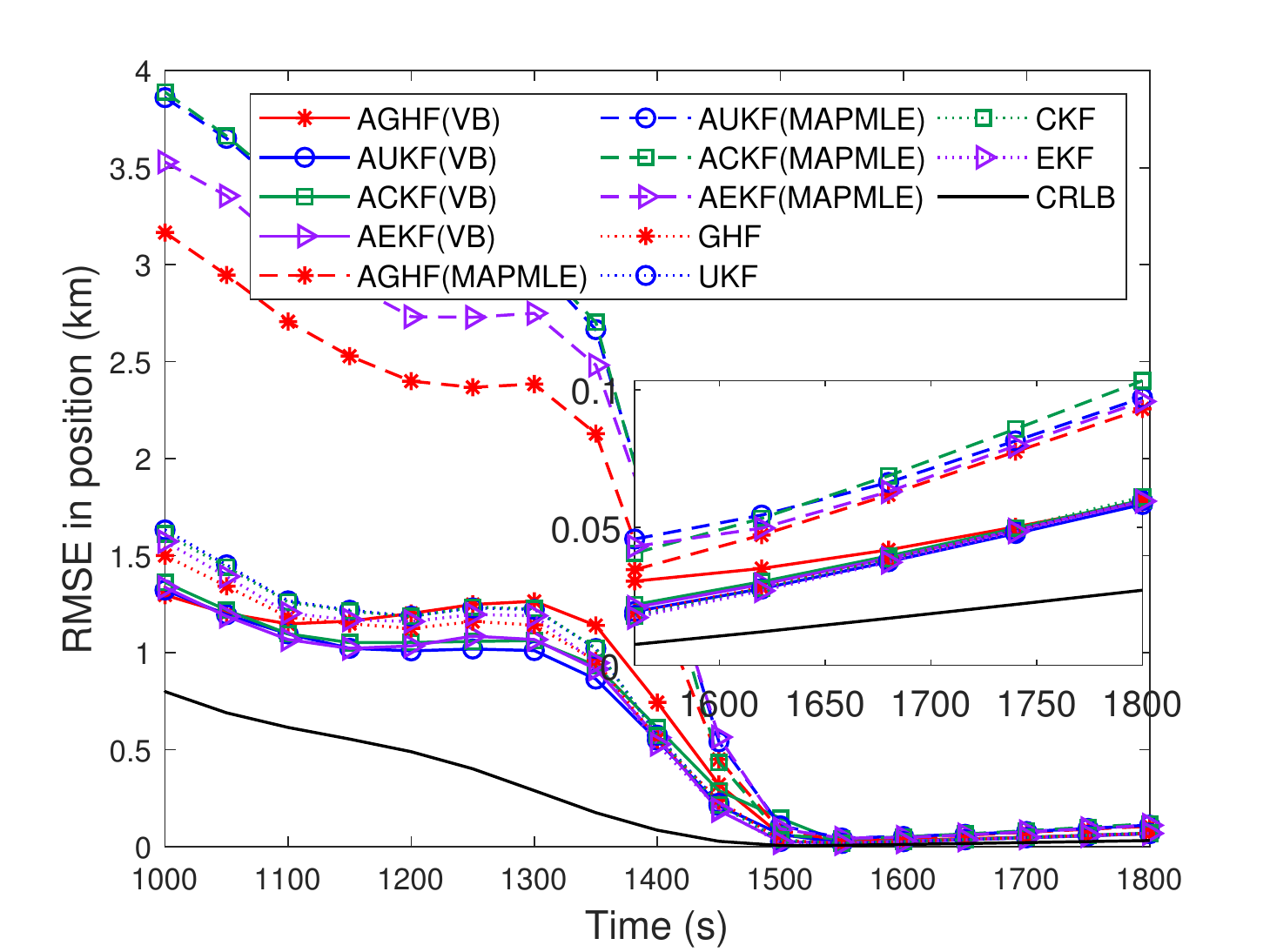}
			\caption{}
			\label{fig_RMSEpstatRSc2}
		\end{subfigure}		
		\begin{subfigure}[b]{0.5\textwidth}
			\centering
			\includegraphics[width=8cm,height=6.5cm]{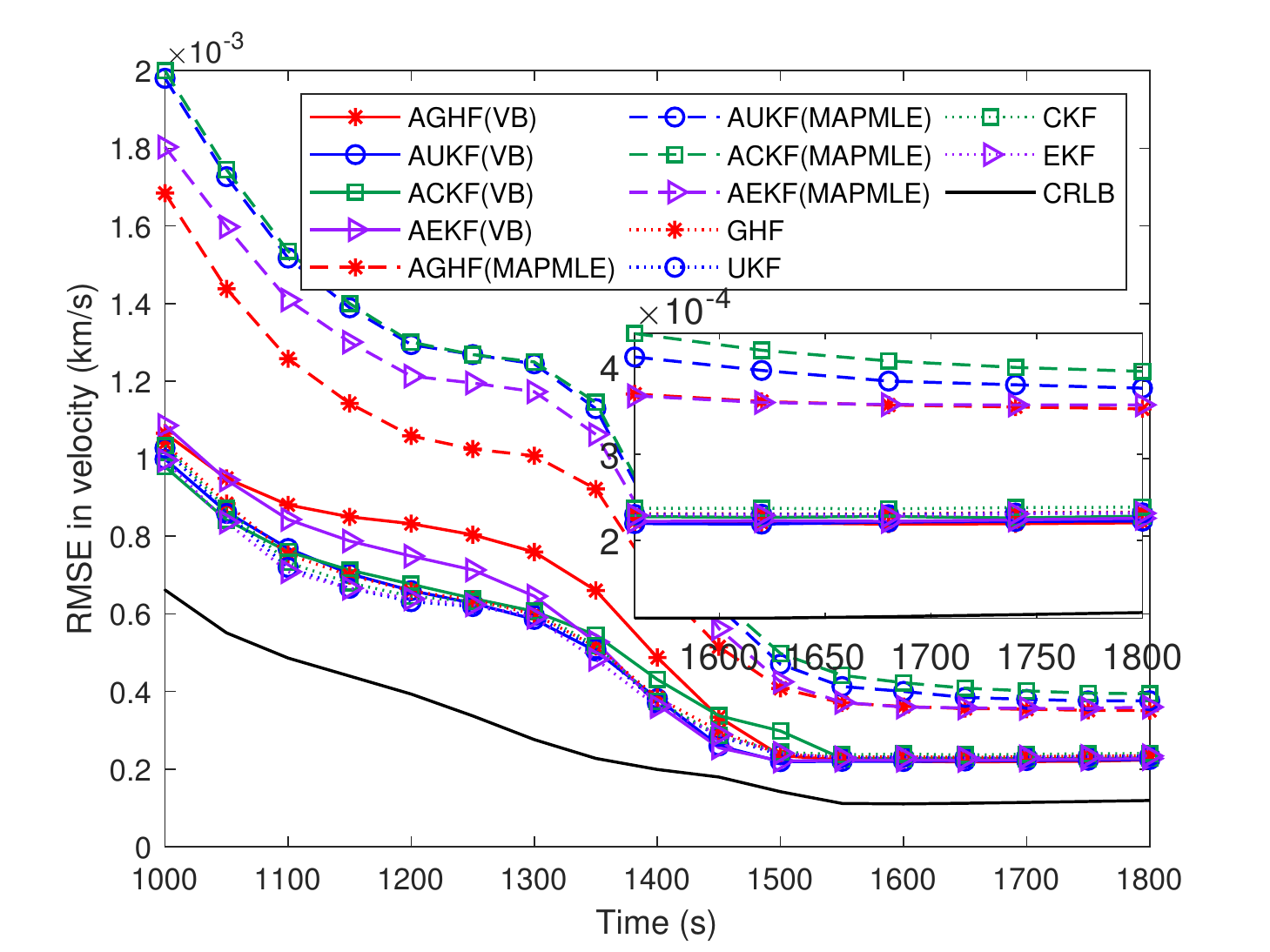}
			\caption{}
			\label{fig_RMSEvstatRSc2}			
		\end{subfigure}
		\caption{RMSE in (a) position, and  (b) velocity for Scenario II, Case I.}
	\end{figure}

	Figure \ref{fig_RMSEpstatRSc1} and \ref{fig_RMSEvstatRSc1} shows the RMSE of position and velocity, respectively for Scenario I and \ref{fig_RMSEpstatRSc2} and \ref{fig_RMSEvstatRSc2} shows RMSE in position and velocity, respectively for Scenario II with fixed value of measurement noise coefficient. All the RMSEs are evaluated excluding the diverged tracks having a track bound of 200 m. From the RMSE plots, we can see that in both scenarios the adaptive filters following the VB approach shows better result than the adaptive filters following the MAPMLE technique. All the adaptive filters have higher RMSE than the nonadaptive filters, where estimation is done using the true values of measurement noise statistics, in Scenario I. In contrast, in Scenario II the adaptive filters with the VB technique show almost similar RMSE with the nonadaptive filters. From the RMSE plots, we can see that the RMSE of the AEKF in the VB approach is lower than the rest of the adaptive filters following the VB approach due to the fact that it has much higher track loss and as the RMSE values are plotted excluding the diverged tracks so the resultant RMSE happens to be less.
	
\begin{figure}
		\begin{subfigure}[b]{0.5\textwidth}
			\centering
			\includegraphics[width=8cm,height=6.5cm]{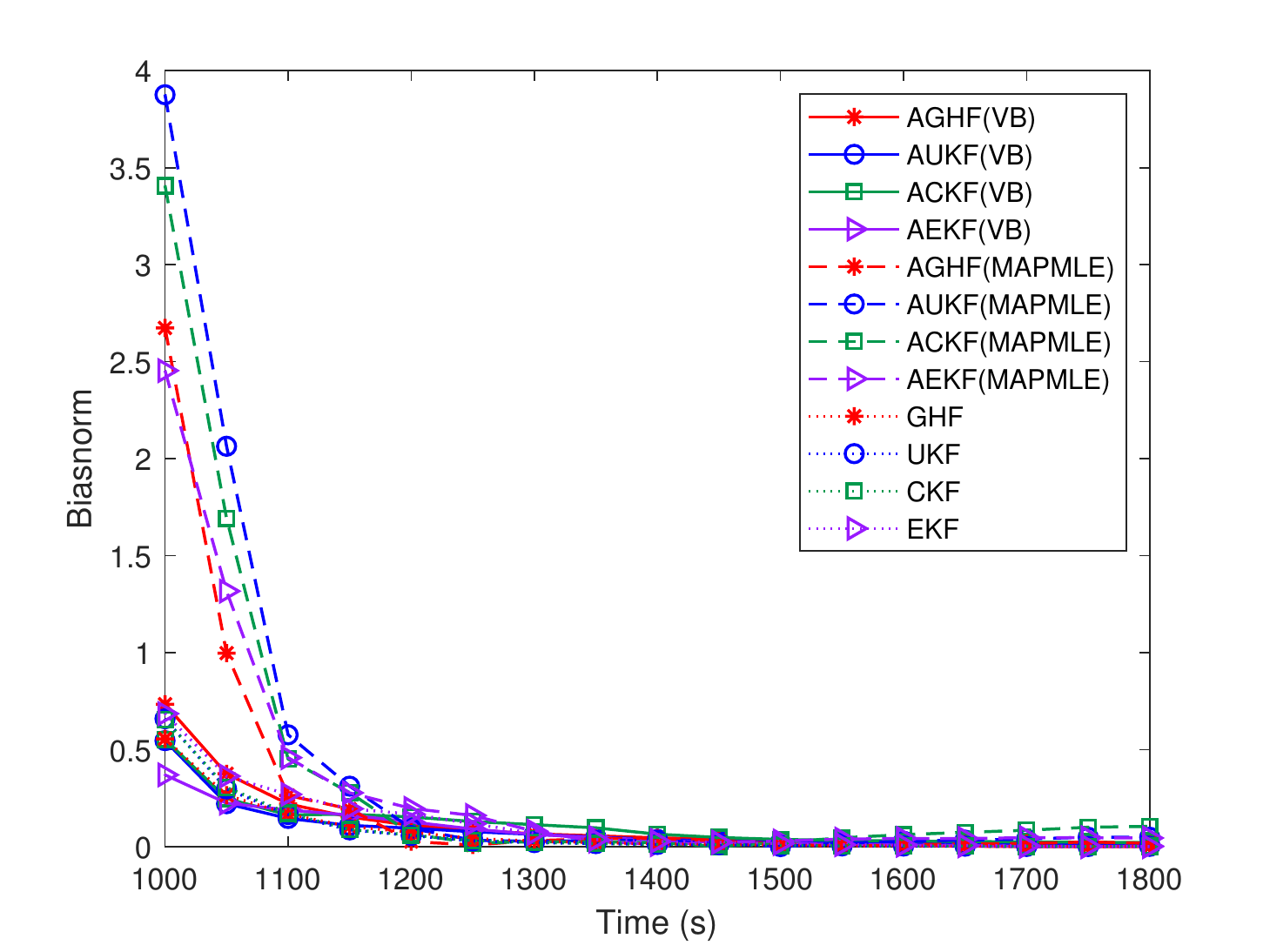}
			\caption{}
			\label{fig_BNstatRSc1}
		\end{subfigure}		
		\begin{subfigure}[b]{0.5\textwidth}
			\centering
			\includegraphics[width=8cm,height=6.5cm]{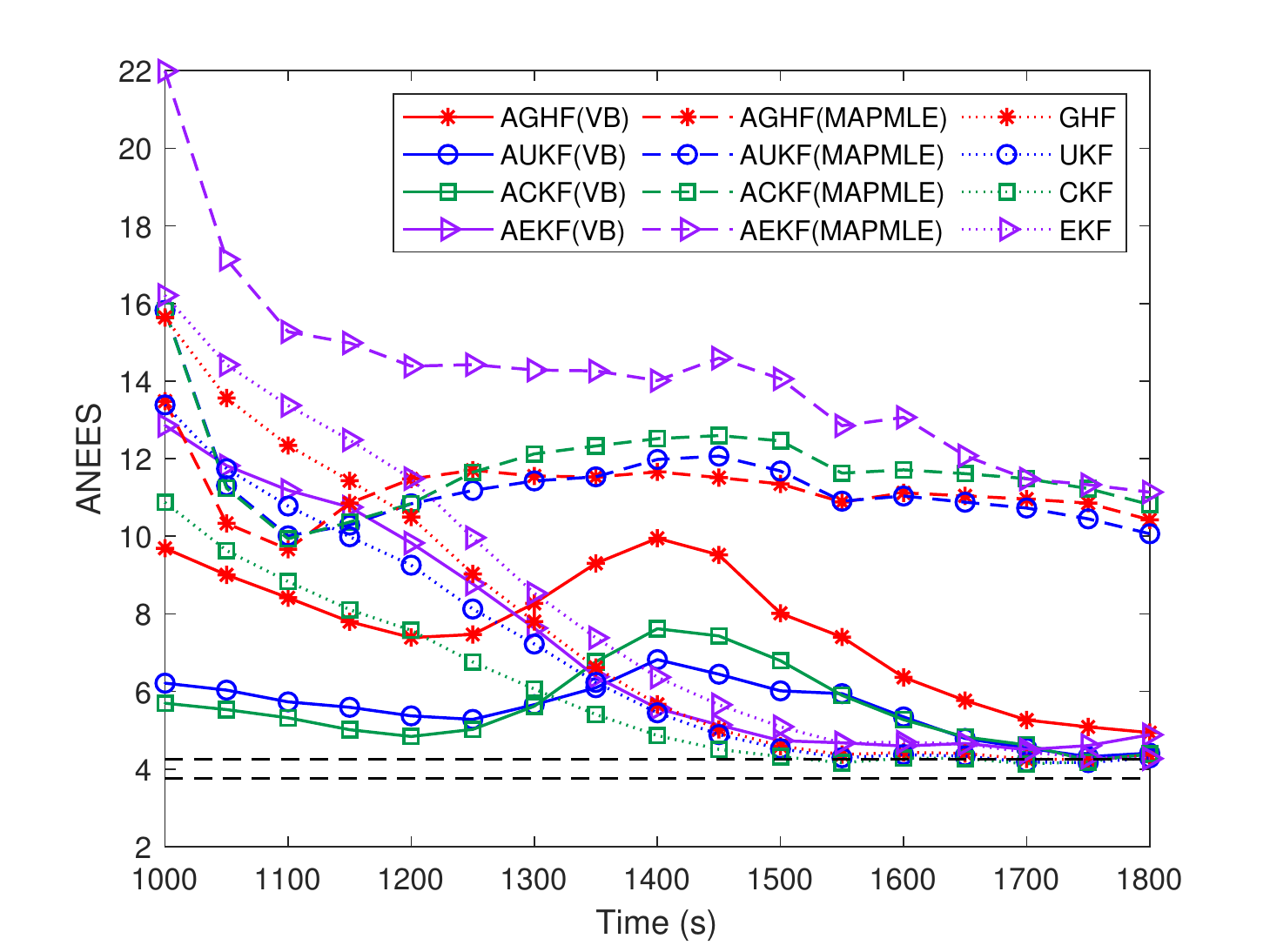}
			\caption{}
			\label{fig_ANEESstatRSc1}			
		\end{subfigure}
		\caption{(a) Bias norm, and  (b) ANEES plots for Scenario I, Case I.}
	\end{figure}

\begin{figure}
		\begin{subfigure}[b]{0.5\textwidth}
			\centering
			\includegraphics[width=8cm,height=6.5cm]{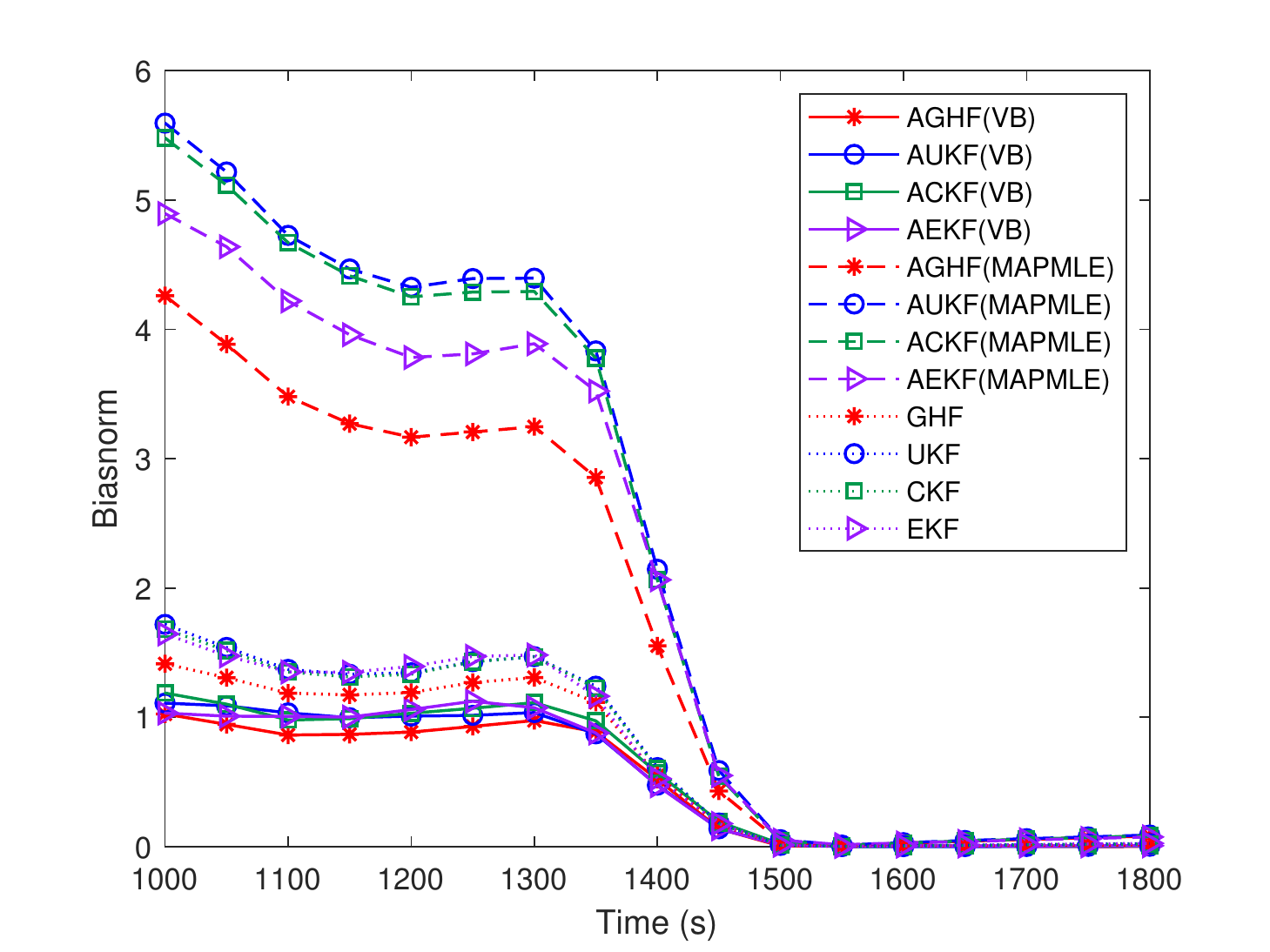}
			\caption{}
			\label{fig_BNstatRSc2}
		\end{subfigure}		
		\begin{subfigure}[b]{0.5\textwidth}
			\centering
			\includegraphics[width=8cm,height=6.5cm]{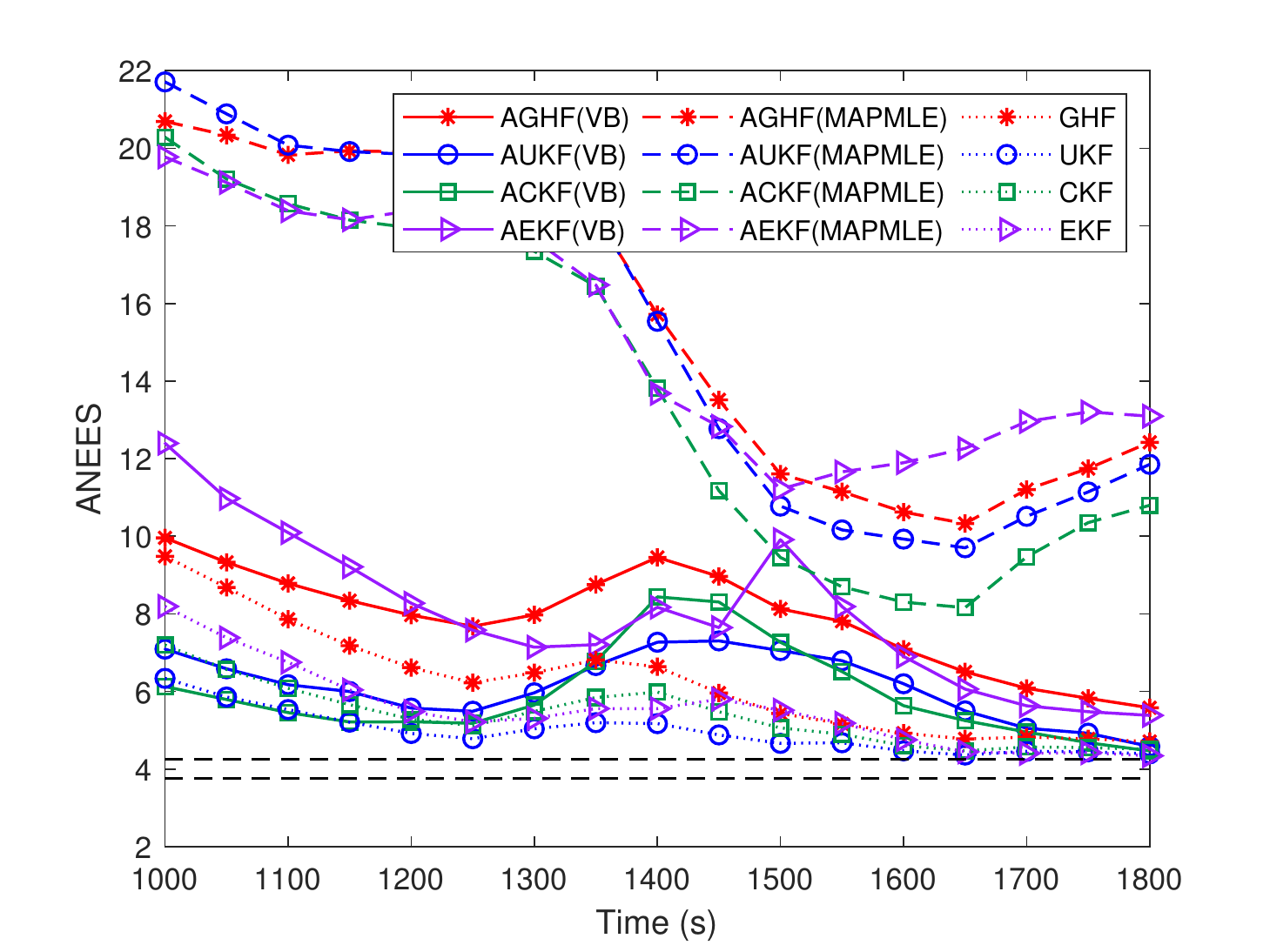}
			\caption{}
			\label{fig_ANEESstatRSc2}			
		\end{subfigure}
		\caption{(a) Bias norm, and  (b) ANEES plots for Scenario II, Case I.}
	\end{figure}

	The bias norm plots for the same are shown in \ref{fig_BNstatRSc1} and \ref{fig_BNstatRSc2} for Scenario I and II, respectively. The bias norm plots are also evaluated excluding the diverged tracks considered with a track bound of 200 m. All the filters' bias norm goes to zero at the end of the simulation signifying zero bias. The ANEES plots for the same are shown in \ref{fig_ANEESstatRSc1} and \ref{fig_ANEESstatRSc2} for Scenario I and II, respectively. The ANEES is plotted excluding the diverged tracks having a track bound of 200 m. The ANEES of the adaptive filters following VB technique is almost similar to the ANIS of the nonadaptive filters at the end of the simulation in both scenarios whereas the adaptive filters following the MAPMLE technique have higher ANEES values than the rest.
	
	The percentage of track loss and the relative execution time for Case I are shown in Table \ref{tracklossCaseI} for both scenarios. The track bound is considered to be 200 m which is used to evaluate the track loss \%. The table also lists the execution time of all the filters relative to the execution time required for the nonadaptive EKF. From the table, we can see that the adaptive filters following the VB approach show much less track loss \% compared to the adaptive filters following the MAPMLE technique at the expense of higher execution time. 
	\begin{table}[h!]
		\renewcommand{\arraystretch}{1.2}
		\setlength{\tabcolsep}{5pt}
		\centering 
		\caption{Percentage of track loss and relative execution time for all the implemented filters for Case I}
		\begin{tabular}{|c|c|c|c|}
			\hline 
			& \multicolumn{2}{c|}{Track loss \%} &  \\ \cline{2-3}
			Filters          &Scenario I&Scenario II&Rel. exe. time \\\hline
			EKF				       &16.35  &19.55  	&1	\\ 	 
			AEKF using VB   & 18.57 	&20.47  	&12.49	\\ 
			AEKF using MAPMLE	   & 40.36 	&	56.13     &1.18 	\\ \hline
			
			CKF					   &1.48  	&4.56 	&2.52	\\ 
			ACKF using VB   	   & 9.48	&17.53	&30.83	\\ 
			ACKF using MAPMLE	   & 34.68 	&52.48       & 1.53	\\ \hline
			
			UKF	                   &1.46  	&4.35&1.08	\\ 
			AUKF using VB  	   & 8.33 		&16.45 	&31.31	\\ 
			AUKF using MAPMLE	   & 34.66 	&49.44       &2.65 	\\ \hline	
			
			GHF	                   &1.44  	&4.10  	&4.48	\\ 
			AGHF using VB  	   &7.43	&15.24		&40.23	\\ 
			AGHF using MAPMLE	   &34.47  		&39.78       &4.23 	\\ \hline
		\end{tabular}
		\label{tracklossCaseI}
	\end{table}

	\subsubsection{Case II: Varying measurement noise covariance}
	
\begin{figure}
		\begin{subfigure}[b]{0.5\textwidth}
			\centering
			\includegraphics[width=8cm,height=6.5cm]{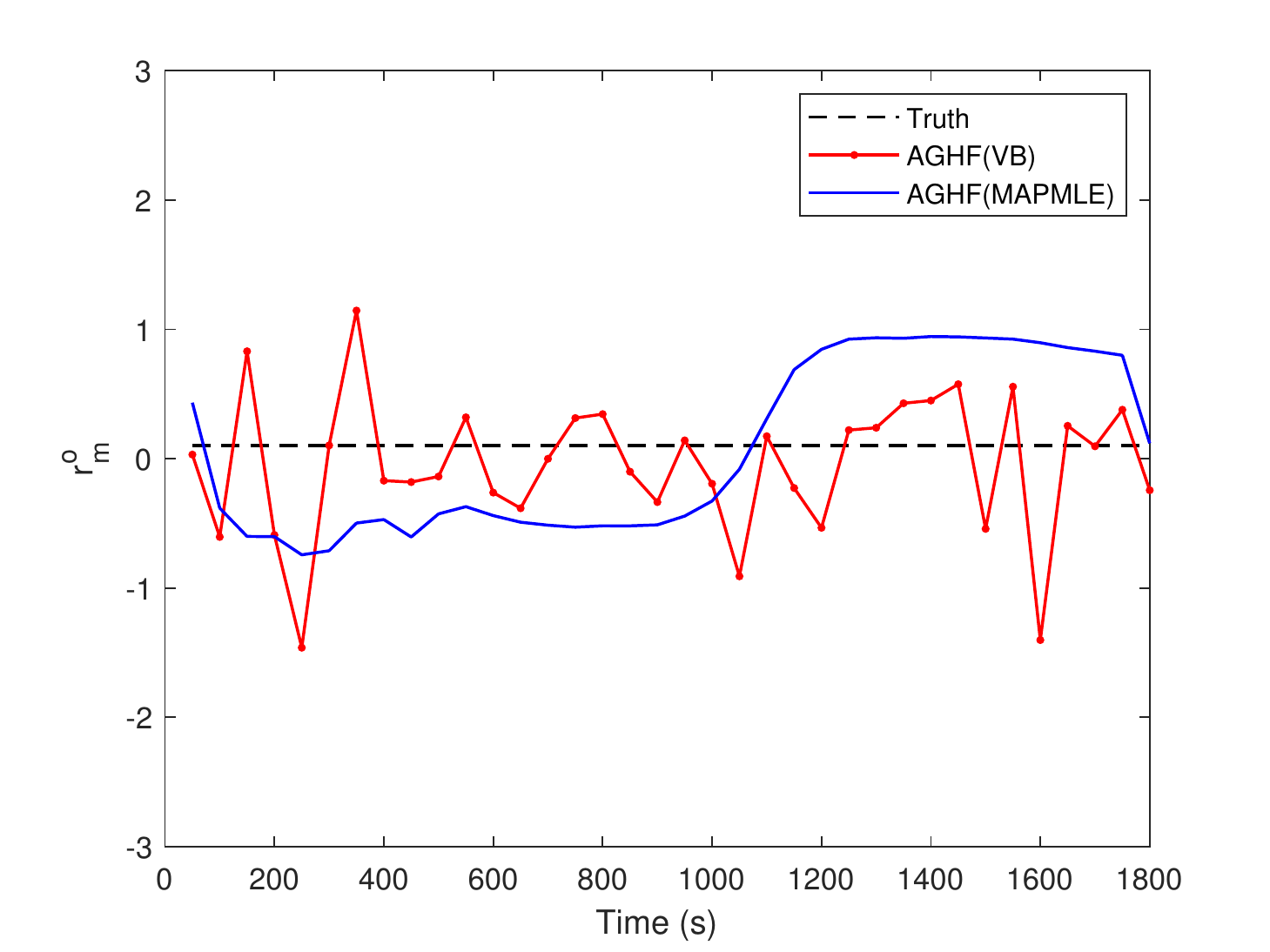}
			\caption{}
			\label{fig_SinglePlotrmSc1}
		\end{subfigure}		
		\begin{subfigure}[b]{0.5\textwidth}
			\centering
			\includegraphics[width=8cm,height=6.5cm]{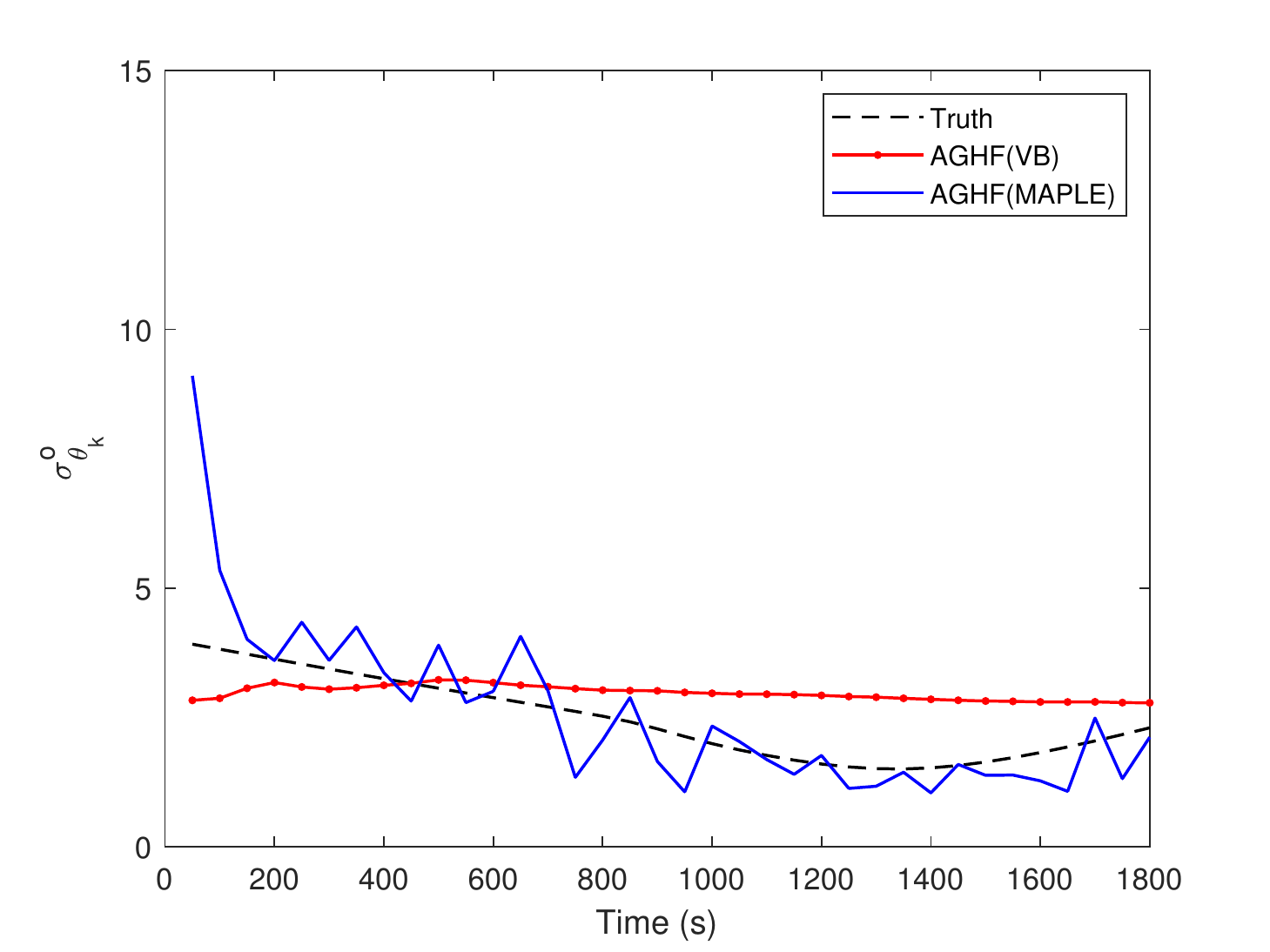}
			\caption{}
			\label{fig_SinglePlotRSc1}			
		\end{subfigure}
		\caption{Estimated (a)$r_m$ and (b) $\sigma_{\theta_k}$ for Scenario I, Case II.}
	\end{figure}

\begin{figure}
		\begin{subfigure}[b]{0.5\textwidth}
			\centering
			\includegraphics[width=8cm,height=6.5cm]{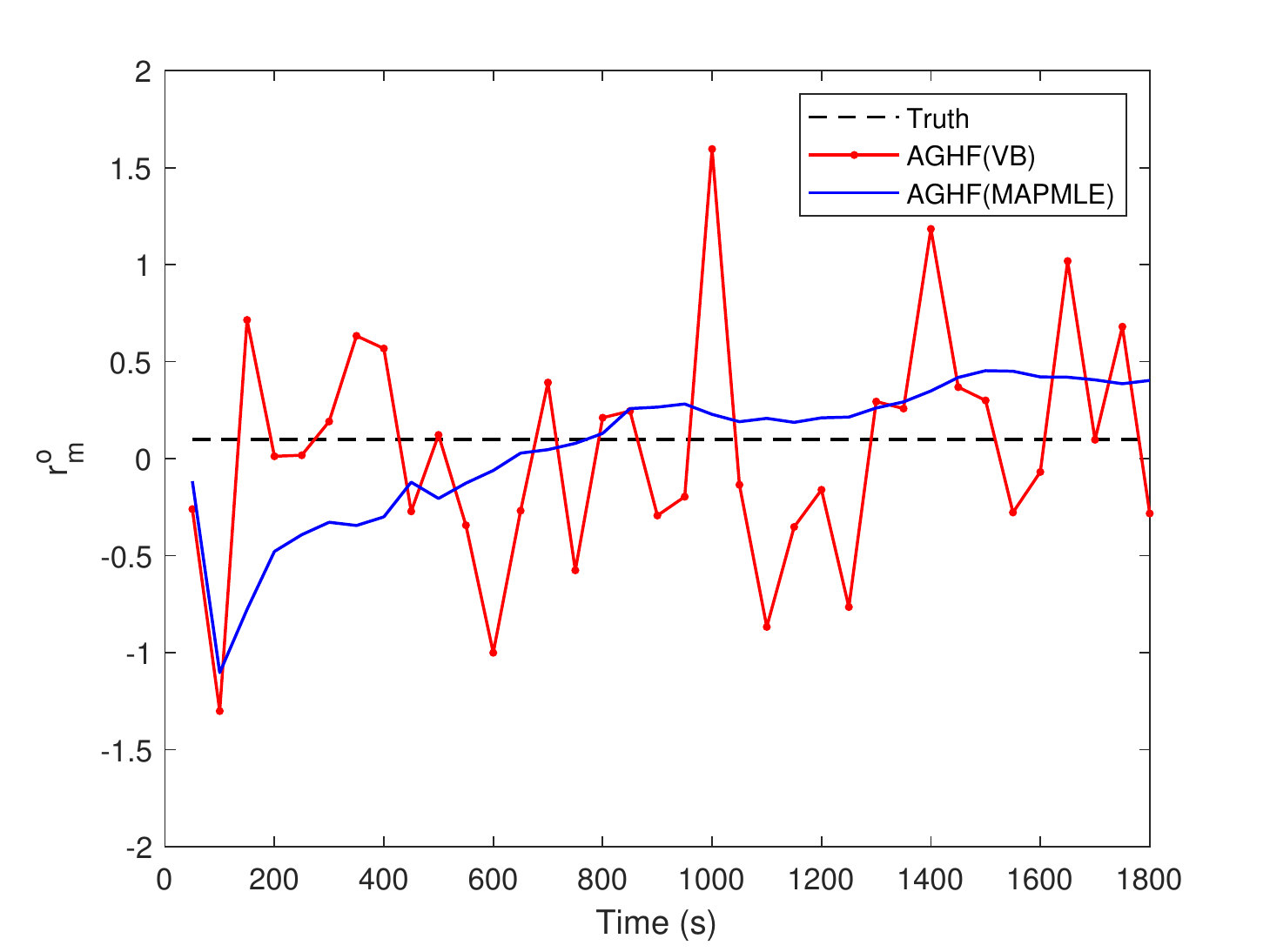}
			\caption{}
			\label{fig_SinglePlotrmSc2}
		\end{subfigure}		
		\begin{subfigure}[b]{0.5\textwidth}
			\centering
			\includegraphics[width=8cm,height=6.5cm]{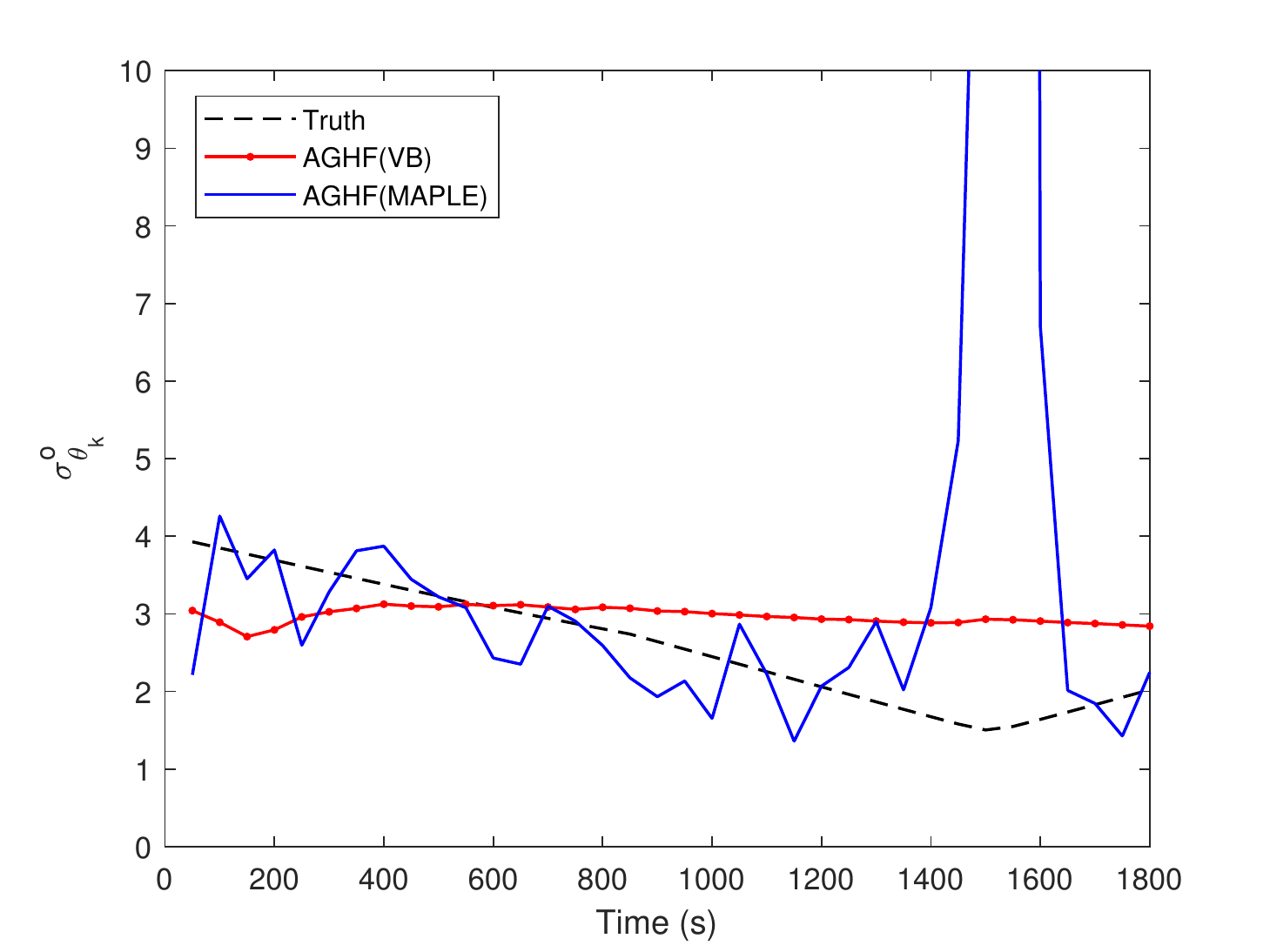}
			\caption{}
			\label{fig_SinglePlotRSc2}			
		\end{subfigure}
		\caption{Estimated (a)$r_m$ and (b) $\sigma_{\theta_k}$ for Scenario II, Case II.}
	\end{figure}

	The single run plots for $r_m$ and $\sigma_{\theta_k}$ estimation for varying measurement noise covariance are shown in Figure \ref{fig_SinglePlotrmSc1} and \ref{fig_SinglePlotRSc1}, respectively for Scenario I and in Figure \ref{fig_SinglePlotrmSc2} and \ref{fig_SinglePlotRSc2}, respectively for Scenario II. We can see that the $r_m$ is estimated in both scenarios and the $\sigma_{\theta_k}$ is also estimated but not too vividly using both adaptive filtering techniques as was in Case I. 
	
\begin{figure}
		\begin{subfigure}[b]{0.5\textwidth}
			\centering
			\includegraphics[width=8cm,height=6.5cm]{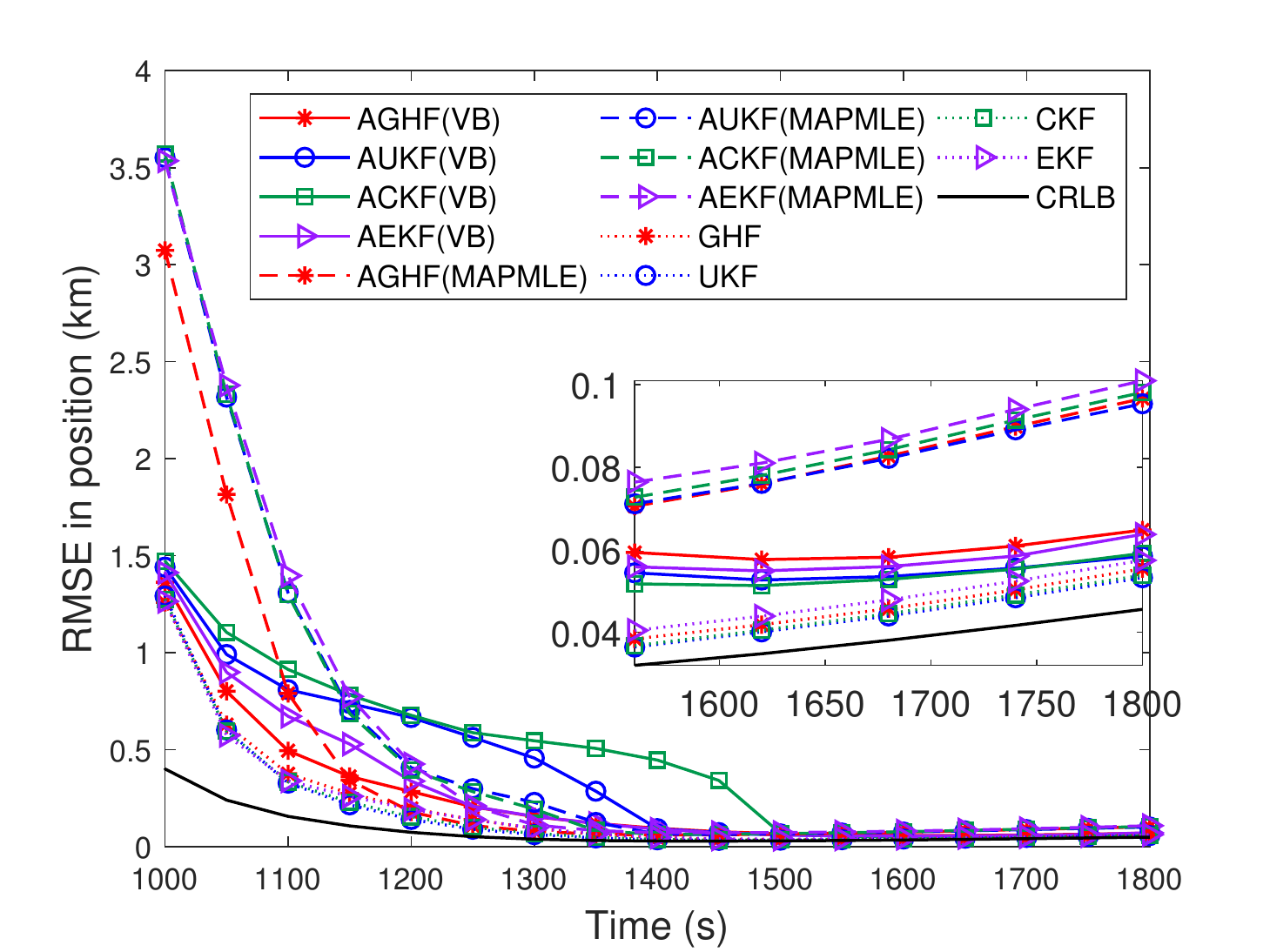}
			\caption{}
			\label{fig_RMSEpVaryRSc1}
		\end{subfigure}		
		\begin{subfigure}[b]{0.5\textwidth}
			\centering
			\includegraphics[width=8cm,height=6.5cm]{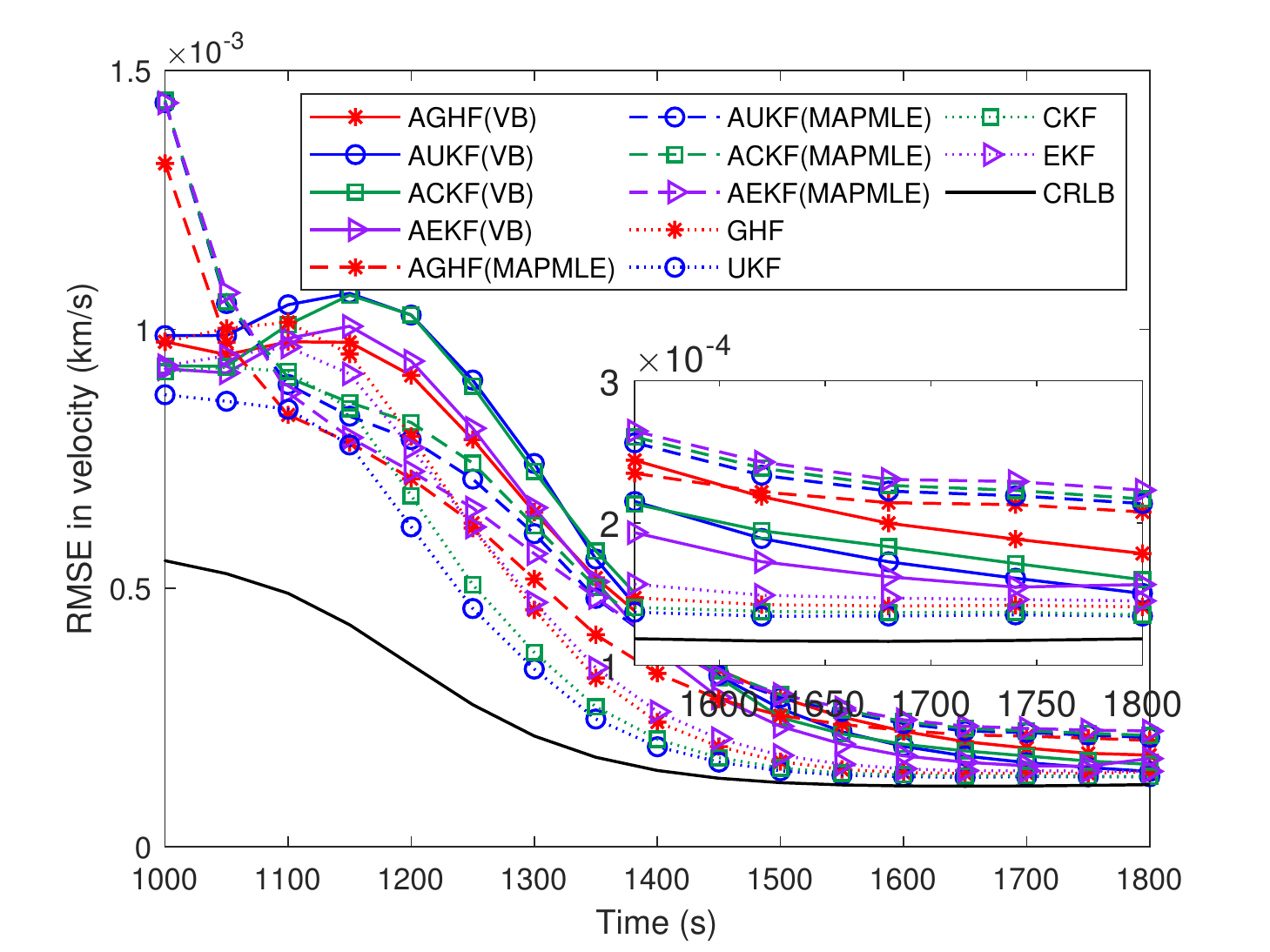}
			\caption{}
			\label{fig_RMSEvVaryRSc1}			
		\end{subfigure}
		\caption{RMSE in (a) position, and  (b) velocity for Scenario I, Case II.}
	\end{figure}

\begin{figure}
		\begin{subfigure}[b]{0.5\textwidth}
			\centering
			\includegraphics[width=8cm,height=6.5cm]{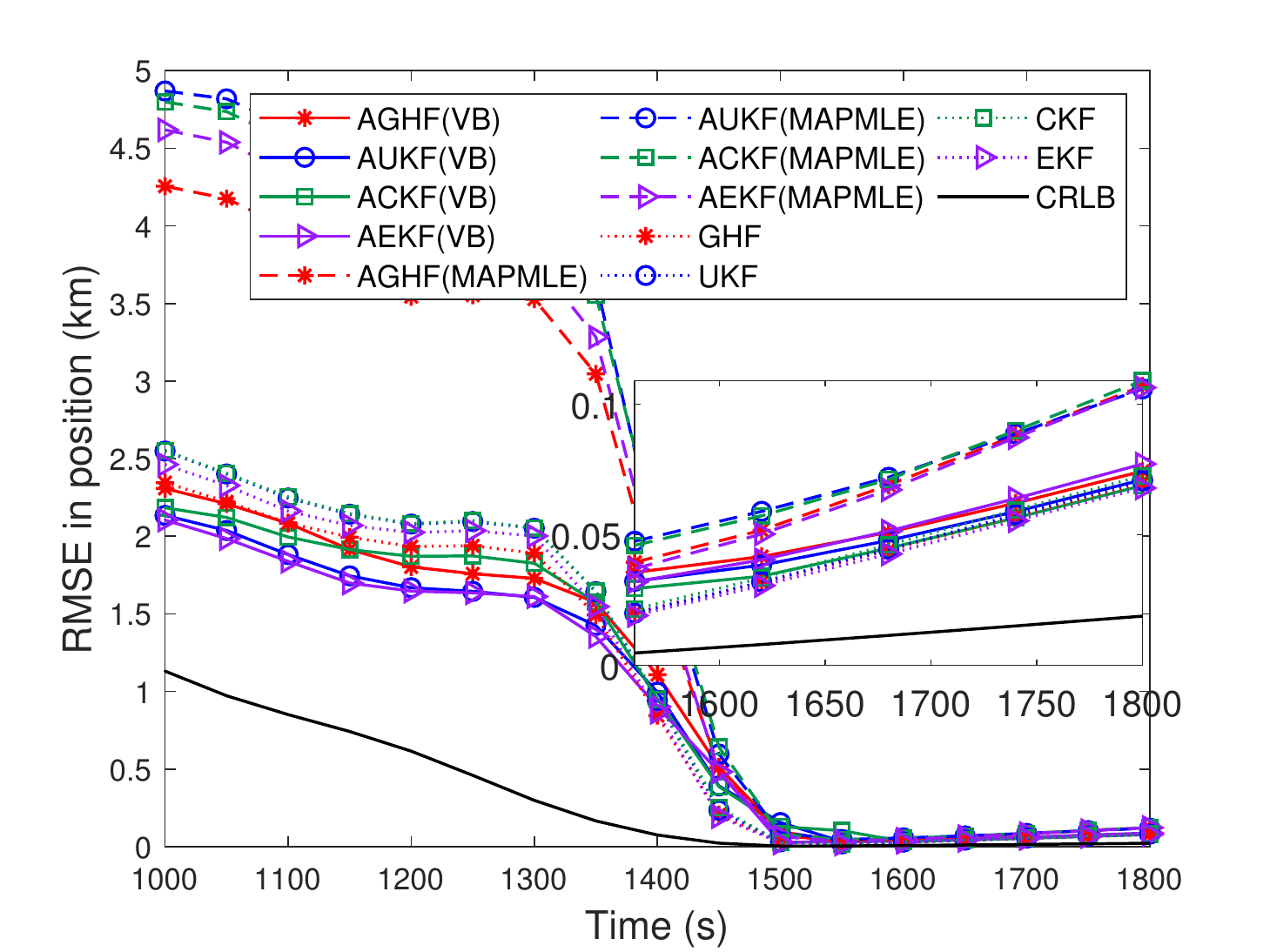}
			\caption{}
			\label{fig_RMSEpVaryRSc2}
		\end{subfigure}		
		\begin{subfigure}[b]{0.5\textwidth}
			\centering
			\includegraphics[width=8cm,height=6.5cm]{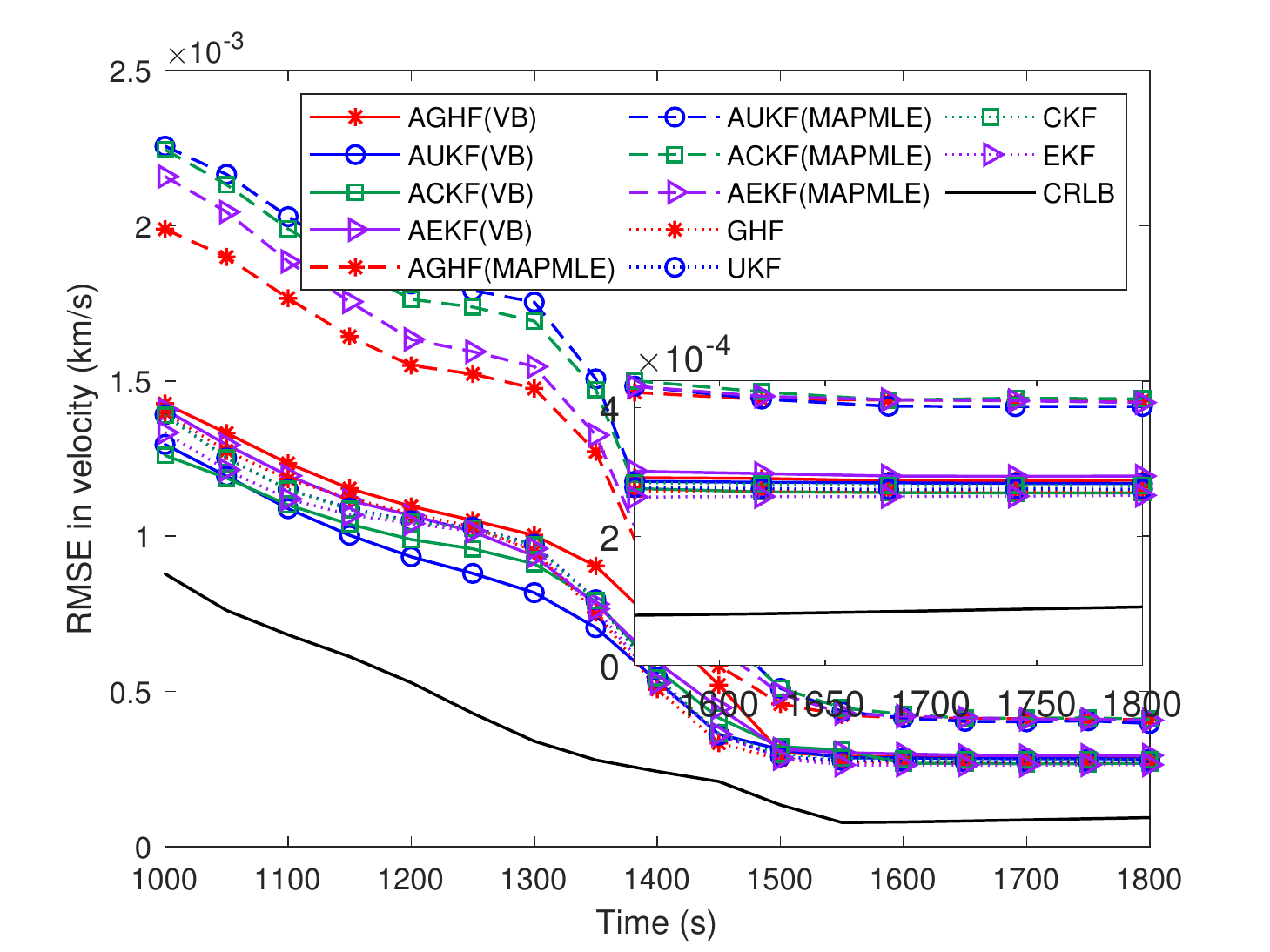}
			\caption{}
			\label{fig_RMSEvVaryRSc2}			
		\end{subfigure}
		\caption{RMSE in (a) position, and  (b) velocity for Scenario II, Case II.}
	\end{figure}

	For Case II, where the measurement noise covariance is considered to be varying the RMSE in position and velocity for Scenario I are shown in \ref{fig_RMSEpVaryRSc1} and \ref{fig_RMSEvVaryRSc1}, respectively, and for Scenario II are shown in \ref{fig_RMSEpVaryRSc2} and \ref{fig_RMSEvVaryRSc2}, respectively. The RMSE is evaluated for 500 Monte Carlo runs excluding the diverged tracking considered with a track bound of 200 m. The RMSE of the adaptive filters following the VB technique is slightly higher than the nonadaptive filters in both scenarios but is lower than the adaptive filters following the MAPMLE technique. In Scenario I, we can see that the RMSE of the AGHF following the VB approach shows higher RMSE than the other adaptive filters following the VB approach because it has a much less percentage of track loss so as the RMSE values are plotted excluding the track loss, the resultant RMSE came to be higher.
	
\begin{figure}
		\begin{subfigure}[b]{0.5\textwidth}
			\centering
			\includegraphics[width=8cm,height=6.5cm]{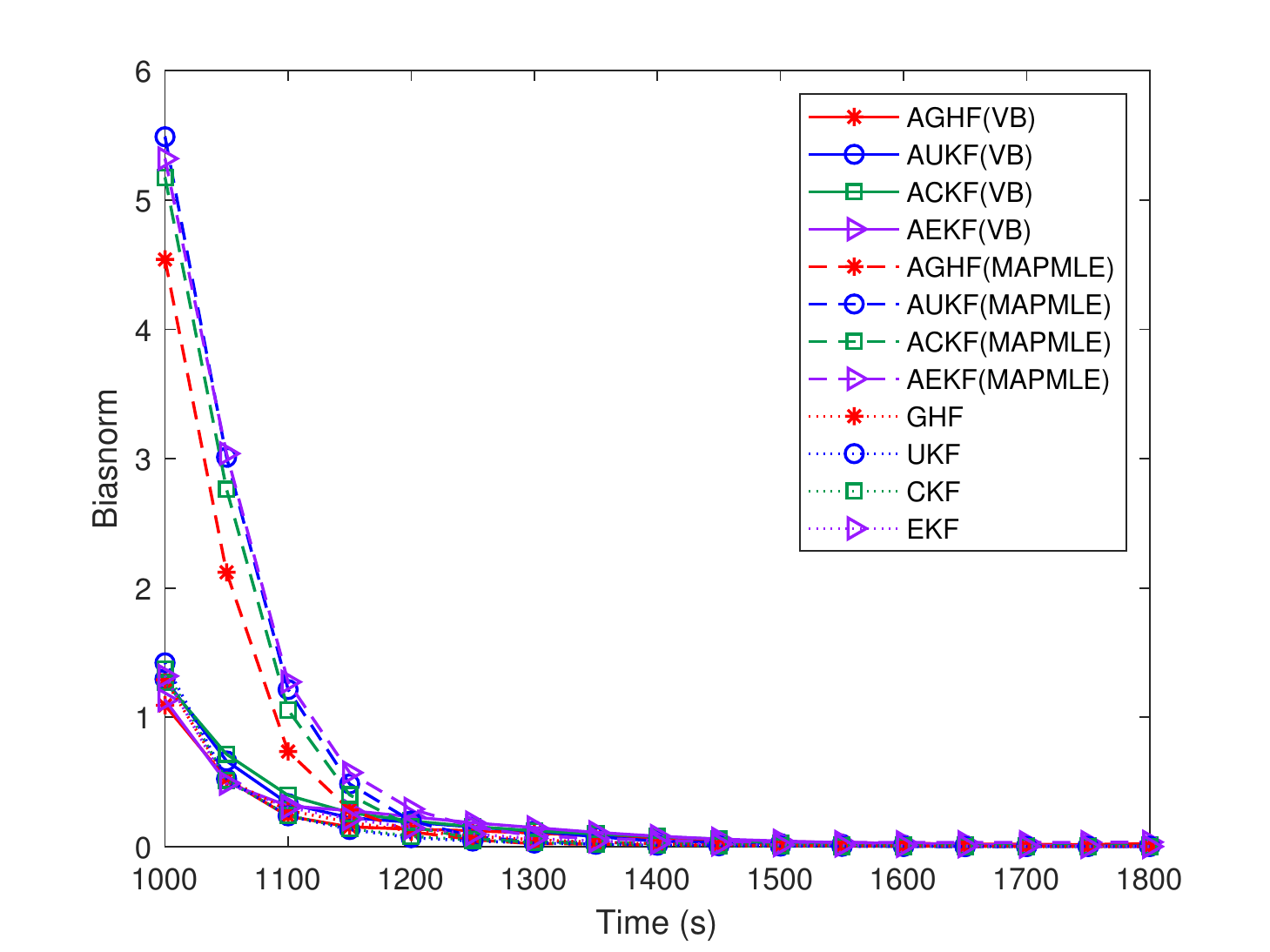}
			\caption{}
			\label{fig_BiasNormVaryRSc1}
		\end{subfigure}		
		\begin{subfigure}[b]{0.5\textwidth}
			\centering
			\includegraphics[width=8cm,height=6.5cm]{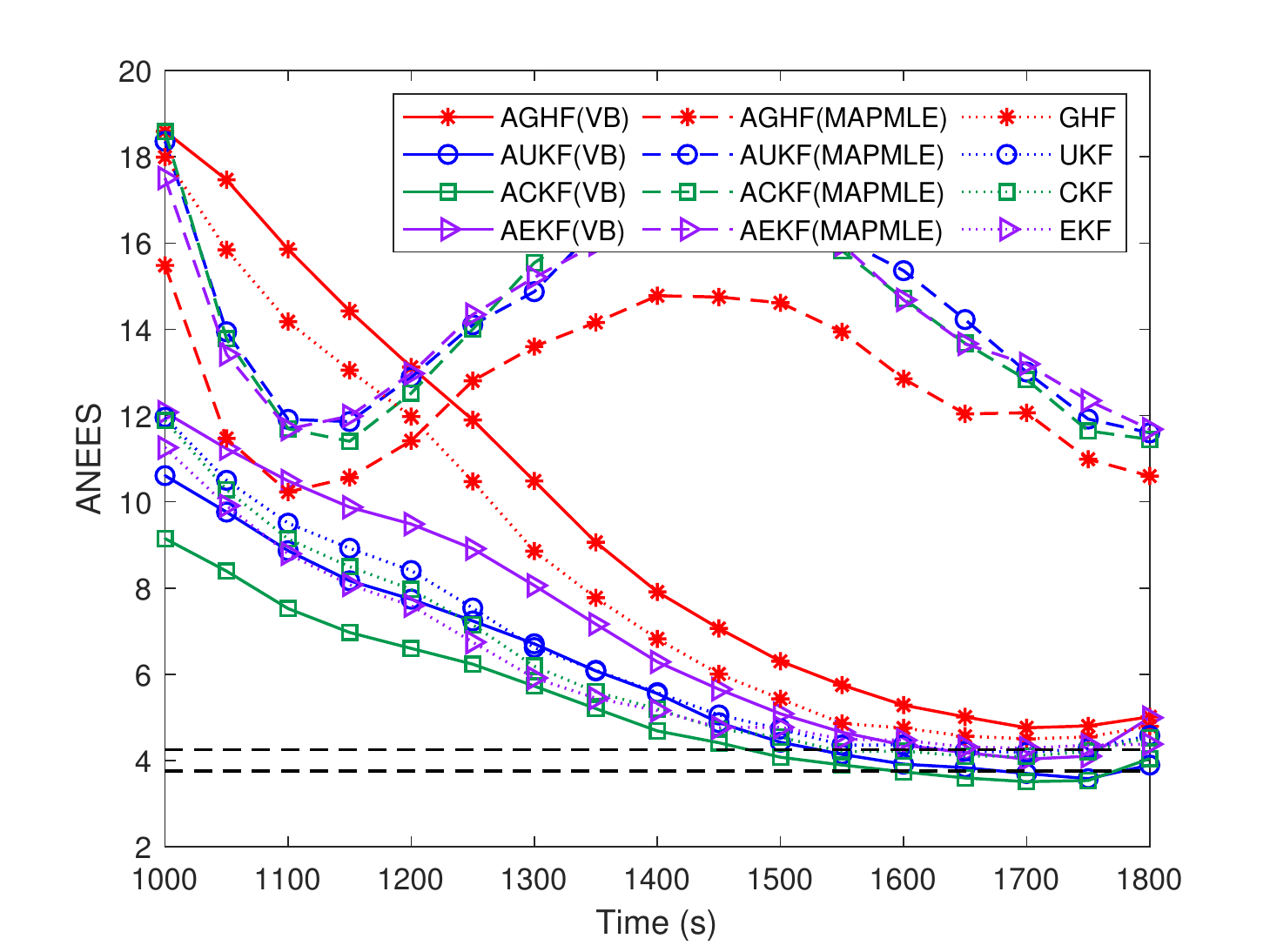}
			\caption{}
			\label{fig_ANEESVaryRSc1}			
		\end{subfigure}
		\caption{(a) Bias norm, and  (b) ANEES plots for Scenario I, Case II.}
	\end{figure}

\begin{figure}
		\begin{subfigure}[b]{0.5\textwidth}
			\centering
			\includegraphics[width=8cm,height=6.5cm]{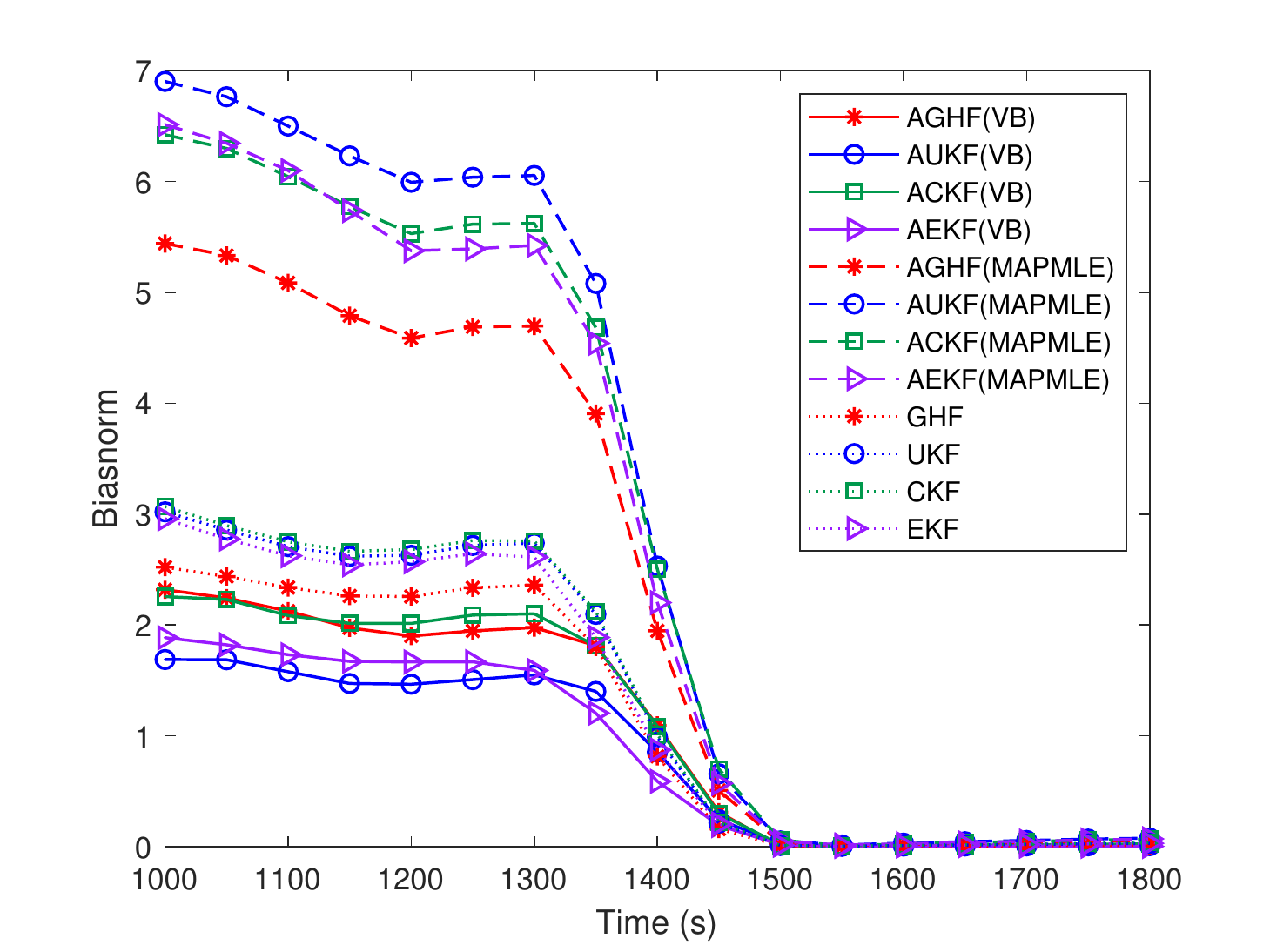}
			\caption{}
			\label{fig_BiasNormVaryRSc2}
		\end{subfigure}		
		\begin{subfigure}[b]{0.5\textwidth}
			\centering
			\includegraphics[width=8cm,height=6.5cm]{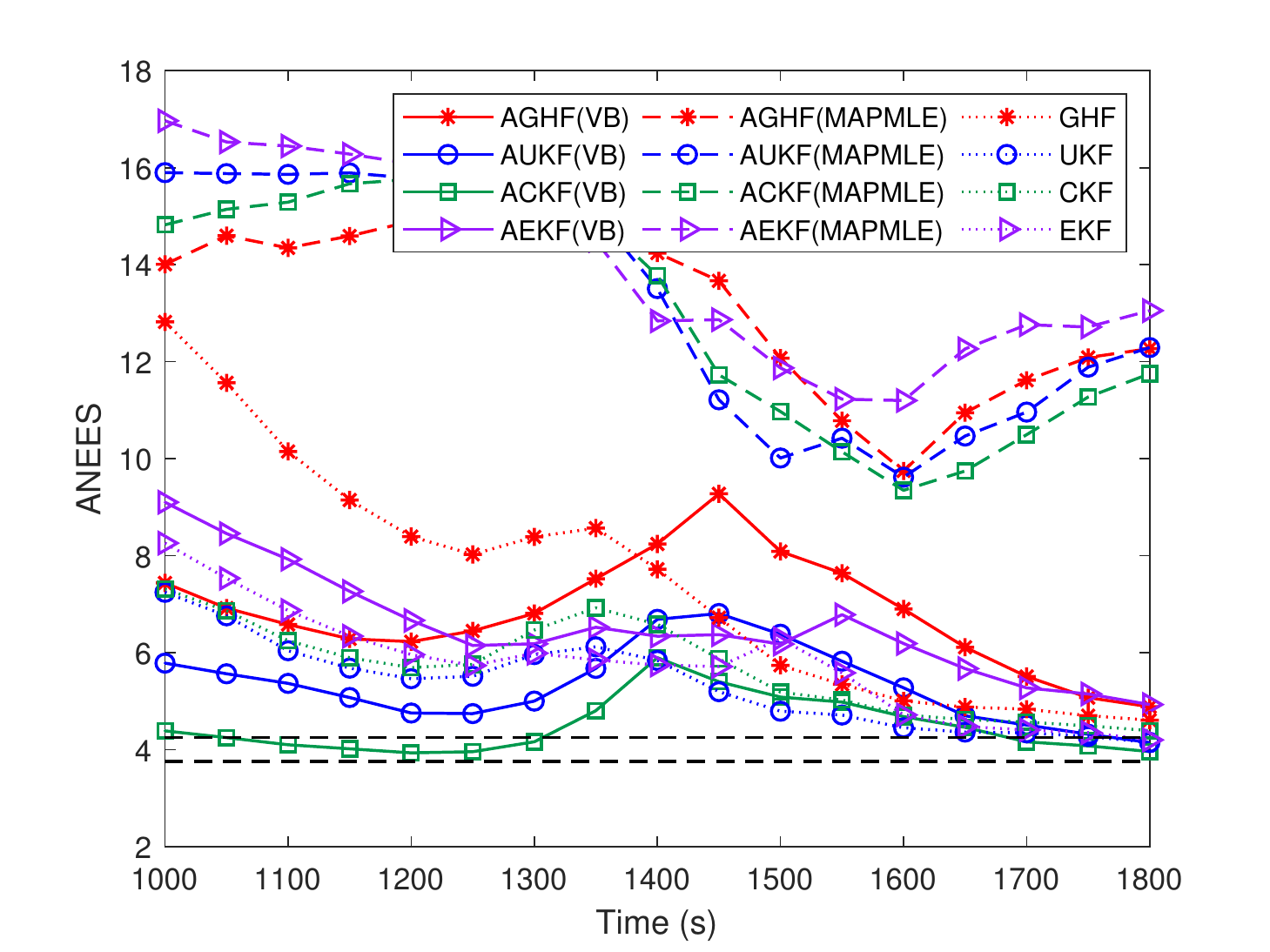}
			\caption{}
			\label{fig_ANEESVaryRSc2}			
		\end{subfigure}
		\caption{(a) Bias norm, and  (b) ANEES plots for Scenario II, Case II.}
	\end{figure}

	The bias norm plot for Scenario I and II considering varying measurement noise covariance is shown in \ref{fig_BiasNormVaryRSc1} and \ref{fig_BiasNormVaryRSc2}, respectively. The biasnorm is also evaluated excluding the diverged tracks considering a track bound of 200 m. In both figures, we can see that the bias norm converges to zero at the end of the simulation signifying zero bias. The ANEES plots for Scenario I and II are shown in \ref{fig_ANEESVaryRSc1} and \ref{fig_ANEESVaryRSc2}, respectively. Here also the ANEES is evaluated excluding the diverged tracks considering a track bound of 200 m. The ANEES plots of the adaptive filters following the VB technique are almost similar to the ANEES of the nonadaptive filters and are also within or around the 95\% probability region shown with the black dashed lines whereas the ANEES of the adaptive filters following the MAPMLE technique are much higher.
	
	The \% of track loss for all the filters for Case II are listed in Table \ref{tracklossCaseII} along with the relative execution time. The track loss \% is evaluated considering a track bound of 200 m. Like Case I, the adaptive filters following the VB approaches show a lower \% of track loss for Case II compared with the adaptive filters following the MAPMLE technique. The relative execution time is evaluated relative to the execution time taken by the nonadaptive EKF filter. The adaptive filters following the VB approach need more time for execution than the adaptive filters following the MAPMLE technique.

	\begin{table}[h!]
		\renewcommand{\arraystretch}{1.2}
		\setlength{\tabcolsep}{5pt}
		\centering 
		\caption{Percentage of track loss and relative execution time for all the implemented filters for Case II}
		\begin{tabular}{|c|c|c|c|}
			\hline 
			& \multicolumn{2}{c|}{Track loss \%} &  \\ \cline{2-3}
			Filters          &Scenario I&Scenario II&Rel. exe. time \\\hline
			EKF				    &13.85  &		21.79    	&1	\\ 
			AEKF using VB  	   &19.58  	&			29.62 &18.31\\ 
			AEKF using MAPMLE	   & 46.85	&	70.97       &1.68\\ \hline
			
			CKF					   & 2.21 	&	11.24    	&1.59	\\ 
			ACKF using VB   	   &9.47  	&	24.32	&31.73\\ 
			ACKF using MAPMLE	   &42.67  	&	63.48	    &2.89 	\\ \hline
			
			UKF	               &  2.13	&	10.29	    	&2.85	\\ 
			AUKF using VB  	   & 8.99 		&23.12	    	&31.79	\\ 
			AUKF using MAPMLE	   &38.53  	&	63.09    &3.12 	\\ \hline	
			
			GHF	               & 2.05 	&	9.44	    	&2.89	\\ 
			AGHF using VB  	   & 5.86 	&	19.83		&47.82	\\ 
			AGHF using MAPMLE	   &37.92		&54.08	  & 4.80	\\ \hline
		\end{tabular}
		\label{tracklossCaseII}
	\end{table}
	
	\section{Discussion and Conclusion}
	This paper presents a variational Bayesian based adaptive filtering framework for the Gaussian approximated filters (GHFs) that can estimate the target trajectory along with the measurement noise statistics. It is based on the assumption that the joint distribution of the measurement noise mean and covariance follow normal inverse Wishart distribution. We tested the proposed technique using traditional GAFs (EKF, CKF, UKF, and GHF) for tracking an underwater target in moderately and highly nonlinear tracking scenarios with bearings-only measurements. 
	Both scenarios are implemented for two cases, one with static measurement noise covariance and another with varying linear covariance with the distance between the target and the ownship.
	We compared the performance of the proposed algorithm with adaptive filtering techniques based on maximum likelihood and maximum aposteriori using RMSE, percentage of track loss, ANEES, bias norm, and computation time. Simulation results showed that the proposed variational Bayesian filter outperformed the adaptive filters based on maximum aposteriori and maximum likelihood.
	\bibliographystyle{ieeetr}
	\bibliography{Bibliography1}

	\appendix
	\section{Proof of Theorem 2:} \label{Proof_Th2}
	\textbf{Appendix A: Proof of Theorem 2}
	
	Let us consider
	\begin{equation}
		p(\mathcal{X}_{k|k-1}|\mathcal{Y}_{1:k})=\mathcal{N}(\mathcal{X}_{k|k-1};\hat{\mathcal{X}}_{k|k-1},P_{k|k-1}),
	\end{equation}
	and
	\begin{equation}\label{Eq_qX}
		\begin{split}
			q(\mathcal{X}_{k|k-1})&=q(\mathcal{Y}_k|\mathcal{X}_{k|k-1})q(\mathcal{X}_{k|k-1}|\mathcal{Y}_{1:k-1})\\&=\mathcal{N}(\mathcal{Y}_k;h(\mathcal{X}_{k|k-1})+r_m,R_k)\mathcal{N}(\mathcal{X}_{k|k-1};\hat{\mathcal{X}}_{k|k-1},P_{k|k-1}).
		\end{split}
	\end{equation}
	
		Using \eqref{Eq_FactorPDF}, considering terms independent of $\mathcal{X}_{k|k-1}$ as constants.
		\begin{equation}
			\begin{split}
				\log p(\mathcal{X}_{k|k-1},R'_k,r'_m,\mathcal{Y}_{1:k})&=\log \dfrac{1}{\sqrt{2\pi R'_k}}-\dfrac{1}{2}(\mathcal{Y}_k-h(\mathcal{X}_{k|k-1})-r'_m)^T{R'}_k^{-1}(\mathcal{Y}_k-h(\mathcal{X}_{k|k-1})-r'_m)+c\\&
				=-\dfrac{1}{2}\log |R'_k|-\dfrac{1}{2}(\mathcal{Y}_k-h(\mathcal{X}_{k|k-1})-r'_m)^T{R'}_k^{-1}(\mathcal{Y}_k-h(\mathcal{X}_{k|k-1})-r'_m)+c_{\mathcal{X}_{k|k-1}}
			\end{split}
		\end{equation}
		Putting $E_{r'_m,R'_k}$ on both side
		\begin{equation}
			\begin{split}
				E_{r'_m,R'_k}\log p(\mathcal{X}_{k|k-1},\mathcal{Y}_{1:k})&=-\dfrac{1}{2}E[\log |R'_k|]-\dfrac{1}{2}(\mathcal{Y}_k-h(\mathcal{X}_{k|k-1})-r'_m)^TE[{R'}_k^{-1}](\mathcal{Y}_k-h(\mathcal{X}_{k|k-1})-r'_m)+c_{\mathcal{X}_{k|k-1}}
			\end{split}
		\end{equation}
		As $\dfrac{1}{2}E[\log |R'_k|]$ is independent of $\mathcal{X}_{k|k-1}$ so they are considered as constant.
		Therefore,
		\begin{equation}
			\begin{split}
				\log q(\mathcal{X}_{k|k-1})&=E_{r'_m,R'_k}\log p(\mathcal{X}_{k|k-1},\mathcal{Y}_{1:k})\\&=-\dfrac{1}{2}(\mathcal{Y}_k-h(\mathcal{X}_{k|k-1})-r'_m)^TE[{R'}_k^{-1}](\mathcal{Y}_k-h(\mathcal{X}_{k|k-1})-r'_m)+c_{\mathcal{X}_{k|k-1}},
			\end{split}
		\end{equation}
		where 
		\begin{equation}\label{Eq_R'}
			E[{R'}_k^{-1}]=\dfrac{\hat{u}'_{k|k-1}+m+1}{\hat{U'}_{k|k-1}},
		\end{equation}

		So, comparing \eqref{Eq_qX} with \eqref{Eq_R'} and considering $\mathcal{N}(\mathcal{X}_{k|k-1},\hat{\mathcal{X}}_{k|k-1},P_{k|k-1})$ as constant we get,
		\begin{equation}\label{Eq_Rk}
			{R}_{k}=(E[{R'}_k^{-1}])^{-1}=\dfrac{\hat{U'}_{k|k-1}}{\hat{u}'_{k|k-1}+m+1}.
		\end{equation}
		\section{Proof of Theorem 3:}\label{Proof_Th3}
		\textbf{Appendix B: Proof of Theorem 3}
		
		Let us consider
		\begin{equation}
			p(r'_m|\mathcal{Y}_{1:k})=\mathcal{N}(r'_m;\mu',\alpha' R'_k),
		\end{equation}
		and
		\begin{equation}
			q(r_m)=\mathcal{N}(r_m;\mu,\alpha R_k).
		\end{equation}
		The terms in \eqref{Eq_FactorPDF} which are independent of $r'_m$ are considered as constant and taking log on both side of the same, we get
		\begin{equation}\label{Eq_LogJointrm}
			\begin{split}
				\log p(\mathcal{X}_{k|k-1},r'_m,R'_k,\mathcal{Y}_{1:k})&=\log \dfrac{1}{\sqrt{2\pi R'_k}}-\dfrac{1}{2}(\mathcal{Y}_k-h(\mathcal{X}_{k|k-1})-r'_m)^T{R'}_k^{-1}(\mathcal{Y}_k-h(\mathcal{X}_{k|k-1})-r'_m)\\&
				+\log \dfrac{1}{\sqrt{2\pi \alpha' R'_k}}-\dfrac{1}{2}(r'_m-\mu')^T(\alpha' R'_k)^{-1}(r'_m-\mu')+c\\&= -\dfrac{1}{2}(\mathcal{Y}_k-h(\mathcal{X}_{k|k-1})-r'_m)^T{R'}_k^{-1}(\mathcal{Y}_k-h(\mathcal{X}_{k|k-1})-r'_m)\\&
				-\dfrac{1}{2}(r'_m-\mu')^T(\alpha' R'_k)^{-1}(r'_m-\mu')+c_{r_m}\\&
			\end{split}
		\end{equation}
		Putting expectation on both sides,
		\begin{equation}\label{Eq_Exprm}
			\begin{split}
				\log q(r_m)&=E_{\mathcal{X}_{k|k-1},R'_k}\log p(\mathcal{X}_{k|k-1},r'_m,R'_k,\mathcal{Y}_{1:k})\\&
				=-\dfrac{1}{2}E[(\mathcal{Y}_k-h(\mathcal{X}_{k|k-1})-r'_m)^T{R'}_k^{-1}(\mathcal{Y}_k-h(\mathcal{X}_{k|k-1})-r'_m)]-\dfrac{1}{2}(r'_m-\mu')^TE[(\alpha' R'_k)^{-1}](r'_m-\mu')+c_{r_m}\\&
				=-\dfrac{1}{2}E[(\mathcal{Y}_k-h(\mathcal{X}_{k|k-1})-r'_m)^T{R'}_k^{-1}(\mathcal{Y}_k-h(\mathcal{X}_{k|k-1})-r'_m)]-\dfrac{1}{2\alpha'}(r'_m-\mu')^TE[{R'}_k^{-1}](r'_m-\mu')+c_{r_m}.
			\end{split}
		\end{equation}
		As $(\mathcal{Y}_k-h(\mathcal{X}_{k|k-1})-r'_m)^T{R'}_k^{-1}(\mathcal{Y}_k-h(\mathcal{X}_{k|k-1})-r'_m)$ is scalar so, 
		\begin{equation}
			tr[(\mathcal{Y}_k-h(\mathcal{X}_{k|k-1})-r'_m)^T{R'}_k^{-1}(\mathcal{Y}_k-h(\mathcal{X}_{k|k-1})-r'_m)]=(\mathcal{Y}_k-h(\mathcal{X}_{k|k-1})-r'_m)^T{R'}_k^{-1}(\mathcal{Y}_k-h(\mathcal{X}_{k|k-1})-r'_m).
		\end{equation}
		Using cyclic property of trace
		\begin{equation}
			tr[(\mathcal{Y}_k-h(\mathcal{X}_{k|k-1})-r'_m)^T{R'}_k^{-1}(\mathcal{Y}_k-h(\mathcal{X}_{k|k-1})-r'_m)]=tr[(\mathcal{Y}_k-h(\mathcal{X}_{k|k-1})-r'_m)(\mathcal{Y}_k-h(\mathcal{X}_{k|k-1})-r'_m)^T{R'}_k^{-1}].
		\end{equation}
		Using this the expectation in the first term of \eqref{Eq_Exprm} can be evaluated as
		\begin{equation}
			\begin{split}
				E[(\mathcal{Y}_k&-h(\mathcal{X}_{k|k-1})-r'_m)^T{R'}_k^{-1}(\mathcal{Y}_k-h(\mathcal{X}_{k|k-1})-r'_m)]=tr\{E[(\mathcal{Y}_k-h(\mathcal{X}_{k|k-1})-r'_m)(\mathcal{Y}_k-h(\mathcal{X}_{k|k-1})-r'_m)^T] E[{R'}_k^{-1}]\}\\&
				=tr\{E[(\mathcal{Y}_k-E[h(\mathcal{X}_{k|k-1})]-r_m+E[h(\mathcal{X}_{k|k-1})]-h(\mathcal{X}_{k|k-1}))(\mathcal{Y}_k-E[h(\mathcal{X}_{k|k-1})]-r_m\\&+E[h(\mathcal{X}_{k|k-1})]-h(\mathcal{X}_{k|k-1}))^T]E[{R'}_k^{-1}]\}\\&
				=tr\{[(\mathcal{Y}_k-E[h(\mathcal{X}_{k|k-1})]-r'_m)(\mathcal{Y}_k-E[h(\mathcal{X}_{k|k-1})]-r'_m)^T-(\mathcal{Y}_k-E[h(\mathcal{X}_{k|k-1})]-r'_m)(E[h(\mathcal{X}_{k|k-1})]]\\&-E[h(\mathcal{X}_{k|k-1})])^T-(E[h(\mathcal{X}_{k|k-1})]]-E[h(\mathcal{X}_{k|k-1})])(\mathcal{Y}_k-E[h(\mathcal{X}_{k|k-1})]-r'_m)^T\\&+(E[h(\mathcal{X}_{k|k-1})]]-E[h(\mathcal{X}_{k|k-1})])(E[h(\mathcal{X}_{k|k-1})]]-E[h(\mathcal{X}_{k|k-1})])^T]E[{R'}_k^{-1}]\}.
			\end{split}
		\end{equation}
		
		Linearizing the measurement function, $(h(\mathcal{X}_{k|k-1}))$ around $(\hat{\mathcal{X}}_{k|k-1})$ following Taylor series, we get $E[h(\mathcal{X}_{k|k-1})]] =  E[h(\mathcal{X}_{k|k-1})]$. Therefore, the above equation becomes 
		\begin{equation}
			E[(\mathcal{Y}_k-h(\mathcal{X}_{k|k-1})-r'_m)^T{R'}_k^{-1}(\mathcal{Y}_k-h(\mathcal{X}_{k|k-1})-r'_m)] =tr\{[(\mathcal{Y}_k-E[h(\mathcal{X}_{k|k-1})]-r'_m)(\mathcal{Y}_k-E[h(\mathcal{X}_{k|k-1})]-r'_m)^T]E[{R'}_k^{-1}]\}.
		\end{equation}
		Again using cyclic property of trace we can rewrite the above equation as
		\begin{equation}
			\begin{split}
				E[(\mathcal{Y}_k&-h(\mathcal{X}_{k|k-1})-r'_m)^T{R'}_k^{-1}(\mathcal{Y}_k-h(\mathcal{X}_{k|k-1})-r'_m)]\\&=tr\{(\mathcal{Y}_k-E[h(\mathcal{X}_{k|k-1})]-r'_m)^TE[{R'}_k^{-1}](\mathcal{Y}_k-E[h(\mathcal{X}_{k|k-1})]-r'_m)\}.
			\end{split}
		\end{equation}
		Again as $(\mathcal{Y}_k-E[h(\mathcal{X}_{k|k-1})]-r'_m)^TE[{R'}_k^{-1}](\mathcal{Y}_k-E[h(\mathcal{X}_{k|k-1})]-r'_m)$ is scalar so we can remove the trace operator and again rewrite the above equation as
		\begin{equation}
			E[(\mathcal{Y}_k-h(\mathcal{X}_{k|k-1})-r'_m)^T{R'}_k^{-1}(\mathcal{Y}_k-h(\mathcal{X}_{k|k-1})-r'_m)]=(\mathcal{Y}_k-E[h(\mathcal{X}_{k|k-1})]-r'_m)^TE[{R'}_k^{-1}](\mathcal{Y}_k-E[h(\mathcal{X}_{k|k-1})]-r'_m).
		\end{equation}
		So,
		\begin{equation}
			\log q(r_m)=(\mathcal{Y}_k-E[h(\mathcal{X}_{k|k-1})]-r'_m)^TE[{R'}_k^{-1}](\mathcal{Y}_k-E[h(\mathcal{X}_{k|k-1})]-r'_m)-\dfrac{1}{2\alpha'}(r'_m-\mu')^TE[{R'}_k^{-1}](r'_m-\mu')+c_{r_m}
		\end{equation}
		As we know the resultant pdf on multiplying 2 Gaussian pdf is also Gaussian in nature \emph{i.e.}, considering 2 nomal distributions of some variable $z_1$ and $z_2$ and $z=z_1\times z_2$ such that, 
		\begin{equation}\label{Eq_Mul2Gauss}
			\mathcal{N}(z_1;m_1,\sigma_1^2)\mathcal{N}(z_2;m_2,\sigma_2^2)=\mathcal{N}(z;\dfrac{m_1\sigma_2^2+m_2\sigma_1^2}{\sigma_1^2+\sigma_2^2},\dfrac{\sigma_1^2\sigma_2^2}{\sigma_1^2+\sigma_2^2})
		\end{equation}
		Using \eqref{Eq_Mul2Gauss} we can say,
		\begin{equation}\label{Eq_muh}
			\begin{split}
				\mu&=\dfrac{\mu'(E[{R'}_k^{-1}])^{-1}+(\mathcal{Y}_k-E[h(\mathcal{X}_{k|k-1})])({\alpha'}^{-1}E[{R'}_k^{-1}])^{-1}}{(E[{R'}_k^{-1}])^{-1}+({\alpha'}^{-1}E[{R'}_k^{-1}])^{-1}}\\&
				=\frac{(E[{R'}_k^{-1}])^{-1}(\mu'+(\mathcal{Y}_k-E[h(\mathcal{X}_{k|k-1})])\alpha')}{(E[{R'}_k^{-1}])^{-1}(1+\alpha')}=\dfrac{\mu'+(\mathcal{Y}_k-E[h(\mathcal{X}_{k|k-1})])\alpha'}{(1+\alpha')},
			\end{split}
		\end{equation}
		where $E[h(\mathcal{X}_{k|k-1})]$ is evaluated using deterministic sample point filtering approach having $N_s$ number of sigma points. $w_j$ is the weight corresponding $j$th sigma point, $\xi_j$. So, 
		\begin{equation}
			E[h(\mathcal{X}_{k|k-1})])=\sum_{j=1}^{N_s}w_j\mathtt{Y}_{j,k|k-1},
		\end{equation}
		where $\mathtt{Y}_{j,k|k-1}=h(\mathtt{X}_{j,k|k-1})=h(S_{k|k-1}\xi_j+\hat{\mathcal{X}}_{k|k-1})$ such that $S_{k|k-1}S_{k|k-1}^T=P_{k|k-1}$.
		Thus, \eqref{Eq_muh} can be rewritten as
		\begin{equation}
			\mu=\dfrac{\mu'+\alpha'(\mathcal{Y}_k-\sum_{j=1}^{N_s}w_j\mathtt{Y}_{j,k|k-1})}{\alpha'+1}.
		\end{equation}
		Using \eqref{Eq_Mul2Gauss} we can also say,
		\begin{equation}
			\begin{split}
				\alpha R_k=\dfrac{(E[{R'}_k^{-1}])^{-1}(E[{R'}_k^{-1}]{\alpha'}^{-1})^{-1}}{(E[{R'}_k^{-1}])^{-1}(1+\alpha')}=\dfrac{(E[{R'}_k^{-1}])^{-1}\alpha'}{1+\alpha'}.
			\end{split}
		\end{equation}
		As, we know from \eqref{Eq_Rk} that, $R_k=(E[{R'}_k^{-1}])^{-1}$, so,
		\begin{equation}
			\alpha=\dfrac{\alpha'}{\alpha'+1}.
		\end{equation}
			\section{Proof of Theorem 4:} \label{Proof_Th4}
			\textbf{Appendix C: Proof of Theorem 4}
			
			Let us consider
			\begin{equation}\label{Eq_pR}
				p(R'_k|\mathcal{Y}_{1:k-1})=IW(R'_k,\hat{u}'_{k|k-1},\hat{U}'_{k|k-1}),
			\end{equation}
			and
			\begin{equation}\label{Eq_qR}
				q(R_k)=IW(R_k,\hat{u}_{k|k-1},\hat{U}_{k|k-1}).
			\end{equation}
			
			
			Considering the terms in \eqref{Eq_FactorPDF} that are independent of $R'_k$ as constant and taking log on both sides we get,
			\begin{equation}
				\begin{split}
					\log p(\mathcal{X}_{k|k-1},r'_m, R'_{k}, \mathcal{Y}_{1:k})&=\log (\dfrac{1}{\sqrt{2\pi(R'_k)}})-\dfrac{1}{2}(\mathcal{Y}_{k}-h({\mathcal{X}}_{k|k-1})-r'_m)^T{R'}_k^{-1}(\mathcal{Y}_{k}-h({\mathcal{X}}_{k|k-1})-r'_m)\\& +\log (\dfrac{1}{\sqrt{2\pi\alpha'R'_k}})-\dfrac{1}{2}(r'_m-\mu')^T(\alpha'R'_k)^{-1}(r'_m-\mu')\\&-\dfrac{\hat{u}'_{k|k-1}+m+1}{2}\log |R'_k|-\dfrac{1}{2}tr(\hat{U}'_{k|k-1}{R'}_{k}^{-1})+c\\&
					=-\dfrac{1}{2}\log|R'_k|+\text{const.}-\dfrac{1}{2}\log|R'_k|+\text{const.}-\dfrac{\hat{u}'_{k|k-1}+m+1}{2}\log |R'_k|\\&-\dfrac{1}{2}tr(\hat{U}'_{k|k-1}{R'}_{k}^{-1})-\dfrac{1}{2\alpha'}(r'_m-\mu')^T(R'_k)^{-1}(r'_m-\mu')\\&-\dfrac{1}{2}(\mathcal{Y}_{k}-h({\mathcal{X}}_{k|k-1})-r'_m)^T{R'}_k^{-1}(\mathcal{Y}_{k}-h({\mathcal{X}}_{k|k-1})-r'_m)\\&
					=-\dfrac{\hat{u}'_{k|k-1}+m+3}{2}\log |R'_k|-\dfrac{1}{2}tr(\hat{U}'_{k|k-1}{R'}_{k}^{-1})\\&-\dfrac{1}{2\alpha'}(r'_m-\mu')^T(R'_k)^{-1}(r'_m-\mu')\\&-\dfrac{1}{2}(\mathcal{Y}_{k}-h({\mathcal{X}}_{k|k-1})-r'_m)^T{R'}_k^{-1}(\mathcal{Y}_{k|k-1}-h({\mathcal{X}}_{k|k-1})-r'_m)+c_{R_k}.
				\end{split}
			\end{equation}
			
			Taking expectation w.r.t. $\mathcal{X}_{k|k-1}$ and $r'_m$ on both sides
			\begin{equation}
				\begin{split}
					E_{\mathcal{X}_{k|k-1},r'_m} \log p(\mathcal{X}_{k|k-1},r'_m, R'_{k}, \mathcal{Y}_{1:k})&= -\dfrac{\hat{u}'_{k|k-1}+m+3}{2}\log |R'_k| -\dfrac{1}{2}tr(\hat{U}'_{k|k-1}{R'}_{k}^{-1})\\&-E_{r'_m}[\dfrac{1}{2\alpha'}(r'_m-\mu')^T(R'_k)^{-1}(r'_m-\mu')]\\&-E_{\mathcal{X}_{k|k-1},r'_m} [\dfrac{1}{2}(\mathcal{Y}_{k}-h({\mathcal{X}}_{k|k-1})-r'_m)^T{R'}_k^{-1}(\mathcal{Y}_{k}-h({\mathcal{X}}_{k|k-1})-r'_m)]+c_{R_k}.\\&
				\end{split}
			\end{equation}
			As, $(\mathcal{Y}_{k}-h({\mathcal{X}}_{k|k-1})-r'_m)^T{R'}_k^{-1}(\mathcal{Y}_{k}-h({\mathcal{X}}_{k|k-1})-r'_m)$ is a scalar, so
			\begin{equation}
				(\mathcal{Y}_{k}-h({\mathcal{X}}_{k|k-1})-r'_m)^T{R'}_k^{-1}(\mathcal{Y}_{k}-h({\mathcal{X}}_{k|k-1})-r'_m)=tr[(\mathcal{Y}_{k}-h({\mathcal{X}}_{k|k-1})-r'_m)^T{R'}_k^{-1}(\mathcal{Y}_{k}-h({\mathcal{X}}_{k|k-1})-r'_m)].
			\end{equation}
			Using cyclic property of trace
			\begin{equation}
				tr[(\mathcal{Y}_{k}-h({\mathcal{X}}_{k|k-1})-r'_m)^T{R'}_k^{-1}(\mathcal{Y}_{k}-h({\mathcal{X}}_{k|k-1})-r'_m)]=tr[(\mathcal{Y}_{k}-h({\mathcal{X}}_{k|k-1})-r'_m)(\mathcal{Y}_{k}-h({\mathcal{X}}_{k|k-1})-r'_m)^T{R'}_k^{-1}].
			\end{equation}
			Also, as $(r'_m-\mu')^T(R'_k)^{-1}(r'_m-\mu')$ is a scalar, so
			\begin{equation}
				(r'_m-\mu')^T(R'_k)^{-1}(r'_m-\mu')=tr[(r'_m-\mu')^T(R'_k)^{-1}(r'_m-\mu')]
			\end{equation}
			Again using cyclic property of trace
			\begin{equation}
				tr[(r'_m-\mu')^T(R'_k)^{-1}(r'_m-\mu')]=tr[(r'_m-\mu')(r'_m-\mu')^T(R'_k)^{-1}].
			\end{equation}
			
			Therefore we can write
			\begin{equation}\label{Eq_CompareR}
				\begin{split}
					E_{\mathcal{X}_{k|k},r'_m} \log p(\mathcal{X}_{k|k-1},r'_m, R'_{k}, \mathcal{Y}_{1:k})&= -\dfrac{\hat{u}'_{k|k-1}+m+3}{2}\log |R'_k| -\dfrac{1}{2}tr(\hat{U}'_{k|k-1}{R'}_{k}^{-1})-\dfrac{1}{2}(B_k+D_k){R'}_k^{-1}+c_{R_k}\\&
					=-\dfrac{\hat{u}'_{k|k-1}+m+3}{2}\log |R'_k| -\dfrac{1}{2}tr[(\hat{U}'_{k|k-1}+B_k+D_k){R'}_{k}^{-1}]+c_{R_k},
				\end{split}
			\end{equation}
			where
			\begin{equation}
				B_k=E_{\mathcal{X}_{k|k-1},r'_m} [(\mathcal{Y}_{k}-h(\mathcal{X}_{k|k-1})-r'_m)(\mathcal{Y}_{k}-h(\mathcal{X}_{k|k-1})-r'_m)^T],
			\end{equation}
			and
			\begin{equation}
				D_k=\dfrac{1}{\alpha'}E_{r'_m}[(r'_m-\mu')(r'_m-\mu')^T].
			\end{equation}

			Comparing the two probabilities of $p$ and $q$ of \eqref{Eq_pR} and \eqref{Eq_qR}, respectively using \eqref{Eq_CompareR} we get,
			\begin{equation}
				\hat{u}_{k|k-1}=\hat{u}'_{k|k-1}+2,
			\end{equation}
			and
			\begin{equation}
				\hat{U}_{k|k-1}=\hat{U}'_{k|k-1}+B_k+D_k.
			\end{equation}

		\end{document}